\RequirePackage{fix-cm}
\documentclass[twocolumn,epjc3]{svjour3}  
\smartqed  % flush right qed marks, e.g. at end of proof
\RequirePackage{graphicx}
\RequirePackage{epstopdf}
\RequirePackage{mathptmx}      % use Times fonts if available on your TeX system
\RequirePackage{flushend}
\RequirePackage[colorlinks,citecolor=blue,urlcolor=blue,linkcolor=blue]{hyperref}
\usepackage[dvipsnames]{xcolor}

\journalname{Eur. Phys. J. A}

\newcommand{\npEMD}{$\delta m^*_{np}$}

\def\eqlab#1{\label{eq:#1}}
\def\eqref#1{Eq.~(\ref{eq:#1})}

\def\figlab#1{\label{fig:#1}}
\def\figref#1{Fig.~(\ref{fig:#1})}

\def\tablab#1{\label{tab:#1}}
\def\tabref#1{Table~(\ref{tab:#1})}

\def\seclab#1{\label{sect:#1}}
\def\secref#1{Section~\ref{sect:#1}}

\begin{document}

\title{In-medium $\Delta(1232)$ potential, pion production in heavy-ion collisions and the symmetry energy%\thanksref{t1}
}


\author{M.D. Cozma\thanksref{addr1,addr2,e1} and M.B. Tsang \thanksref{addr1} }

%\author{M.B. Tsang\thanksref{addr1}}

%\thankstext{e1}{e-mail: fauthor@example.com}
\thankstext{e1}{\emph{Email address:} dan.cozma@theory.nipne.ro}%

\institute{National Superconducting Cyclotron Laboratory, Michigan State University, East Lansing, MI 48824, USA \label{addr1}
           \and
           Department of Theoretical Physics, IFIN-HH, Reactorului 30, 077125 M\v{a}gurele-Bucharest, Romania \label{addr2}}
%\institute{Department of Theoretical Physics, IFIN-HH, Reactorului 30, 077125 M\v{a}gurele-Bucharest, Romania \label{addr2}}

\date{Received: date / Accepted: date}

\maketitle

\begin{abstract}
Using the dcQMD transport model, the isoscalar and isovector in-medium potentials of
the $\Delta$(1232) baryon are studied and information regarding their effective strength is obtained from a comparison to experimental pion production data in heavy-ion collisions below 800 MeV/nucleon impact energy. The best description is achieved for an isoscalar potential moderately more attractive than the nucleon optical potential and a rather small isoscalar relative effective mass m$^*_\Delta \approx$ 0.45. For the isovector component only a constraint between the potential's strength at saturation and the isovector effective mass difference can be extracted, which depends on quantities such as the slope of the symmetry energy and the neutron-proton effective mass difference. These results are incompatible with the usual assumption, in transport models, that the $\Delta$(1232) and nucleon potentials are equal. The density dependence of symmetry energy can be studied using the high transverse momentum tail of pion multiplicity ratio spectra. Results are however correlated with the value of neutron-proton effective mass difference. This region of spectra is shown to be affected by uncertain model ingredients such as the pion potential or in-medium correction to inelastic scattering cross-sections at levels smaller than 10$\%$. Extraction of precise constraints for the density dependence of symmetry energy above saturation will require experimental data for pion production in heavy-ion collisions below 800 MeV/nucleon impact energy and experimental values for the high transverse momentum tail of pion multiplicity ratio spectra accurate to better than 5$\%$.

\end{abstract}

%===============================================
\section{Introduction}
\label{intro}

The isospin dependent part of the equation of state of nuclear matter (asy-EoS), commonly known as the symmetry energy (SE) remains among the most debated topics in nuclear physics. Its relevance for the structure of rare isotopes, dynamics of heavy-ion collisions and properties of neutron stars and associated phenomena has been long recognized and has prompted numerous experimental and theoretical studies ~\cite{Li:2008gp,Lattimer:2006xb,Baldo:2016jhp,Lattimer:2015nhk}. By combining results for various experimental observables with phenomenological models~\cite{Chen:2005ti,Trippa:2008gr,Tsang:2012se,Brown:2013mga,Zhang:2013wna,Danielewicz:2013upa,Morfouace:2019jky} and theoretical many-body simulations of nuclear matter \cite{Kruger:2013kua,Drischler:2016djf,Drischler:2015eba} a consistent description of SE at sub-saturation densities has been achieved. 

The recent observation of a binary neutron star merger by the LIGO-VIRGO collaboration \cite{TheLIGOScientific:2017qsa,Abbott:2018exr} has opened up the possibility of studying the asy-EoS in the vicinity of twice saturation density (2$\rho_0$) by means of correlations between tidal polarizability of neutron stars ($\Lambda$), their radii and ultimately symmetry energy \cite{Lattimer:2015nhk,Fattoyev:2017jql}. However, a unique correspondence between $\Lambda$ and the SE does not exist, due to a  degeneracy of the sensitivity to the slope ($L$) and curvature ($K_{sym}$) parameters of the asy-EoS around 2$\rho_0$ \cite{Zhang:2018vbw}. Nuclear physics laboratory experiments, astrophysical observations and theoretical studies are thus needed to provide lacking complementary information. More recently, developments of theoretical many-body calculations based on chiral effective interactions have made predictions of the asy-EoS up to 2$\rho_0$ with unprecedented accuracy possible \cite{Drischler:2020hwi}, calling for independent confirmation of these results.
 
Heavy-ion collisions (HIC) provide an unique opportunity to study nuclear matter at densities exceeding $\rho_0$ in the laboratory. To this end several promising observables have been identified: the ratio of neutron-to-proton yields of squeezed out nucleons~\cite{Yong:2007tx}, charged pion multiplicity ratio (PMR) and its spectral ratio \cite{Li:2004cq,Hong:2013yva}, elliptic flow related observables~\cite{Li:2002qx} and others. Using neutron-to-proton and neutron-to-charged particles elliptic flow ratios compatible constraints for the value of $L$ have been extracted using different transport models \cite{Russotto:2011hq,Wang:2014rva,Russotto:2016ucm,Cozma:2017bre}. Extrapolations to 2$\rho_0$ are still uncertain due to limited experimental accuracy and suboptimal average density probed by these observables in AuAu collisions at 400 MeV/nucleon impact energy. 

The charged pion multiplicity ratio has attracted considerable attention from the community. Reaching at a consistent picture for the density dependence of SE has been however elusive up to this moment \cite{Xiao:2009zza,Feng:2009am,Xie:2013np,Hong:2013yva,Song:2015hua,Cozma:2016qej}. Numerous studies have attempted to remedy the problem, but have only succeeded in unvealing the sensitivity of PMR to additional model ingredients \cite{Song:2015hua,Cozma:2016qej,Cozma:2014yna,Zhang:2017mps,Zhang:2017nck,Zhang:2018ool,Ikeno:2016xpr,Ikeno:2019mne,Cui:2019dmk}. In recent years, the Transport Model Evaluation Project (TMEP) has aimed towards understanding differences between existing models and formulating benchmark calculations that every realistic model should reproduce \cite{Xu:2016lue,Zhang:2017esm,Ono:2019ndq}. The model used in this study is part of that effort.

In Refs. \cite{Cozma:2016qej,Cozma:2014yna} a Quantum Molecular Dynamics (QMD) model has been employed in an attempt to explain the FOPI experimental pion production data \cite{Reisdorf:2010aa} by inclusion of threshold effects \cite{Ferrini:2005jw,Ferini:2006je} that arise as a consequence of imposing total energy conservation of the system. This requirement is often not properly treated in semi-classical transport models, in spite of its relevance for the existence of thermodynamic equilibrium \cite{Zhang:2017nck}. The crucial ingredients for the computation of threshold effects are the in-medium potential energy of nucleons, resonances (only $\Delta$(1232) close to the vacuum production threshold) and pions.

The knowledge of the isoscalar $\Delta$(1232) potential (ISDP) is uncertain, with empirical information contradicting microscopical calculations \cite{O'Connell:1990zg,Bodek:2020wbk,Hirata:1977hg,Horikawa:1980cv,Oset:1987re,GarciaRecio:1989xa,deJong:1992wm,Baldo:1994fk}. Discrepancies among results of microscopical models have also been noted and are often related to details of how pion-nucleon and pion-nucleon-delta couplings have been extracted from few-body experimental data. In particular, including (or omitting) processes such as $N\Delta \rightarrow N\Delta$, $N\Delta \rightarrow \Delta\Delta$, $NN \rightarrow \Delta\Delta$ and $\Delta\Delta \rightarrow \Delta\Delta$ in models used to describe nucleon-nucleon scattering data was proven to have an impact on the determined strength of the $\Delta$ potential \cite{Baldo:1994fk}. No information is avaiblable about the isovector component of the $\Delta$(1232) potential (IVDP). These quantities are also relevant for determining the threshold density above which $\Delta$(1232) occurs in neutron stars, with impact on the maximum mass of such objects \cite{Drago:2014oja,Cai:2015xga,Zhu:2016mtc,Kolomeitsev:2016ptu,Li:2019tjx} and in the analysis of neutrino physics experimental data \cite{Bodek:2020wbk}.

In view of the above, it is customary to set, in transport models, the $\Delta$(1232) potential (DPOT) in terms of that of nucleons using a simple Ansatz based on the decay channels of this resonance into nucleon-pion pairs \cite{Li:2002yda}. The significance of this assumption was recognized and a large sensitivity of PMR to the magnitude of these potentials was evidenced in Ref. \cite{Cozma:2014yna}. Subsequently, it was shown that the density dependence of the SE can be studied by using PMR supplemented by the ratio of average transverse momenta of charged pions \cite{Cozma:2016qej}. The latter observable is needed in order to constrain the strength of IVDP, which was varied using a scaling parameter. In that study the ISDP was kept fixed, equal to that of the nucleon, in spite of previously proven dependence of PMR on its strength \cite{Cozma:2014yna}.

Extracting the asy-EoS from low and intermediate energy regime experiments is further complicated by uncertainties stemming from the rather poorly constrained momentum/energy dependence of nuclear interactions, usually quantified in terms of effective masses \cite{Morfouace:2019jky,Li:2013ola,Li:2014qta,Zhang:2015qdp,Zhang:2017hvh,Li:2018lpy} and the degeneracy of effects induced by the isoscalar mass, the neutron-proton effective mass difference (\npEMD) and the density dependence of SE on observables \cite{Li:2018lpy,Kong:2017nil,Malik:2018juj}.

The present study builds on the results of Refs. \cite{Cozma:2016qej,Cozma:2014yna}. The goal is to describe all pionic observables, not just ratios of multiplicities or average transverse momenta, in an attempt to reduce residual model dependence originating from the isoscalar part of the interaction. To achieve this goal the DPOT is treated as an independent quantity. For both isoscalar and isovector components freedom is built into parametrizations as to allow independent assigning of potential depths at saturation and effective masses. Details of the transport model, parametrizations used for DPOT and benchmarking calculations for nucleonic observables are presented in \secref{2}. The observables relevant for constraining of DPOT parameters and their extraction from experimental data are described in \secref{3}. In \secref{4} the feasibility of constraining the density dependence of SE from pionic observables is reassessed, together with a study of the impact of other relevant model parameters, such as \npEMD. A section devoted to summary and conclusions follows.

%Your text comes here. Separate text sections with
%===============================================
\section{The model}
\seclab{2}
%Text with citations \cite{RefB} and \cite{RefJ}.
%----------------------------------------------
\subsection{Transport model}
\seclab{2a}

Quantum molecular dynamics transport models provide a semi-classical framework for theoretical description
of heavy ion reactions by accounting for relevant quantum aspects such as stochastic scattering and Pauli blocking of nucleons. They deliver a solution for the time dependence of the density matrix of the system by the
method of the Weyl transformation applied to the many-body Schr\"odinger equation. 
Generally, the expectation values for the position and momentum operators can be shown to satisfy the classical
Hamiltonian equations of motion ~\cite{deGroot:1972aa,Hartnack:1997ez}.
These can be factorized to each particle by approximating the total wave-function of the system
as the product of individual nucleon wave functions, represented by Gaussian wave packets of
finite spread in phase space,
\begin{eqnarray}
\eqlab{eomqmd}
\frac{d\vec{r}_i}{dt}=\frac{\partial \langle U_i \rangle}{\partial \vec{p}_i}+\frac{\vec{p}_i}{m},\qquad
\frac{d\vec{p}_i}{dt}=-\frac{\partial \langle U_i \rangle}{\partial \vec{r}_i}\,.
\end{eqnarray}
The average of the potential operator is understood to be taken over the entire phase-space
and weighted by the Wigner distribution of particle $i$. The potential operator $U_i$ is in
this case the sum of the Coulomb and strong interaction potential operators. 

In the present study a variant developed over the last couple of years, dubbed dcQMD, is used ~\cite{Cozma:2017bre,Cozma:2016qej,Cozma:2014yna}. It traces its origin to the T\"ubingen QMD model transport model developed in the 90's and early 2000's ~\cite{Khoa:1992zz,UmaMaheswari:1997ig,Fuchs:2000kp,Shekhter:2003xd}.

In the present model, the relativistic relation between mass, energy and momentum is used in all kinematic equations. Consequently the kinetic term in \eqref{eomqmd} is replaced by its relativistic counterpart. To be complete, the effective classical Hamiltonian reads
\begin{eqnarray}
\eqlab{hamiltonian}
 H&=&\sum_i \sqrt{p_i^2+m_i^2}+\sum_{i,j,j>i}\,\bigg[\frac{A_u+A_l}{2}+\tilde\tau_i\,\tilde\tau_j\,\frac{A_l-A_u}{2}\bigg]\,u_{ij} \\
 &+&\sum_{i,j,j>i}\bigg[(C_l+C_u)+\tilde\tau_i\,\tilde\tau_j\,(C_l-C_u)\bigg]\frac{u_{ij}}{1+(\vec{p_i}-\vec{p_j})^2/\Lambda^2}\nonumber\\
 &+&\sum_i\,\frac{B}{\sigma +1}\,[1-x\tilde\tau_i\,\beta_i\,]\,u_i^\sigma+\frac{D}{3}\,[1-y\tilde\tau_i\,\beta_i]\,u_i^2 
 +\sum_{i,j,j>i} U_{ij}^{Coul} \nonumber
\end{eqnarray}
where $\tilde\tau_i$=-$\tau_i/T_i$, $u_{ij}=\rho_{ij}/\rho_0$ is the partial relative interaction density of particles $i$ and $j$ with $u_i=\sum_{j\neq i} u_{ij}$ and $\beta_i$ is the isospin asymmetry at the location of particle $i$. Here
$T_i$ and $\tau_i$ denote the isospin and isospin projection of particle $i$ respectively. It is straightforward to show that the momentum independent part of the interaction leads to the expression of the energy per particle presented in
\eqref{eos} up to symmetry potentials of second and higher order. The momentum dependent term above represents a finite
particle number approximation to the corresponding expression in \eqref{eos}. 

The scattering term includes elastic and inelastic two-baryon collisions ($N+N\rightarrow N+N$, $N+N\rightarrow N+R$, $N+R\rightarrow N+R'$, etc.), resonance decays into a pion-nucleon or pion-resonance pairs ($R\rightarrow N+\pi$ and $R\rightarrow R'+\pi$) and single pion absorption reactions ($\pi+N\rightarrow R$). Collision processes that consist of 3-particle initial or final states (as for example non-resonant background pion production $N+N \rightarrow N+N+\pi$) have not been considered. Non-resonant pion production contributions are needed at invariant masses close to the production threshold to describe experimental data \cite{Engel:1996ic,Shyam:1996id,Effenberger:1996im}. Their inclusion in the scattering term is, in the context of using the geometrical Bertsch prescription for collision validation \cite{Bertsch:1988ik} and requirement of conservation of total energy of the system \cite{Cozma:2014yna}, technically challenging, leading to a significant slow down of computations, and has thus not been attempted.

The vacuum Li-Machleidt \cite{Li:1993rwa,Li:1993ef} and Cugnon ${\it et\;al.}$ \cite{Cugnon:1980rb} parametrizations of elastic nucleon-nucleon cross-sections are used below and above pion production threshold respectively. They are modified in nuclear matter using an empirical factor depending on density and relative momentum, but not isospin asymmetry. Such a modification has been found necessary to describe stopping and flow observables at low and intermediate energy heavy-ion collision~\cite{Barker:2016hqv,Basrak:2016cbo,Wang:2013wca,Li:2011zzp}. The FU3FP4 parametrization in Ref. \cite{Li:2011zzp} has been found to lead to the best description of stopping and flow, see \secref{2c}. For this choice, elastic cross-sections are multiplied by a factor depending on the local density $\rho$ and relative momentum $p$ of the scattering nucleons
\begin{eqnarray}
F(\rho,p)=\left\{\begin{array}{ll}
1 & \textrm{if {\it p} $>$ 1.0 GeV/c}\\
\frac{F_\rho -1}{1+(p/p_0)^\kappa}+1 & \textrm{if {\it p} $\leq$ 1.0 GeV/c}
\end{array} \right. \\
\mathrm{with}\qquad F_\rho=\lambda+(1-\lambda)\,Exp[-\frac{\rho}{\zeta\,\rho_0}]\,.\nonumber
\end{eqnarray}
The parameters in the above expression take the following values: $p_0$=0.30 GeV/c, $\kappa$=8, $\lambda$=1/6
and $\zeta$=1/3. Theoretically computed medium-modified cross-sections \cite{Li:1993rwa,Li:1993ef} fail to lead to a good description of stopping at low impact energies.

The Huber ${\it et\,\,al.}$ parametrizations for vacuum inelastic nucleon-nucleon cross sections \cite{Huber:1994ee} are used. 
They lead to charged pion production cross-section that underpredict experimental values for $nn$/$pp$  and $np$ reactions by 20$\%$
and 40$\%$ respectively, at an impact energy of 400 MeV/nucleon. The discrepancy can be alleviated by including non-resonant background contributions. Charged pions emitted in HIC originate predominantly from $nn$/$pp$ 
collisions since for these channels production cross-sections are an order of magnitude larger than in $np$ reactions. Consequently, explicit non-resonant terms to pion production multiplicities can be neglected at the impact energies of interest for this study, as their omission can, as a first approximation, be compensated by modifying the strength of the $\Delta$(1232) potentials (see \secref{3a}). This approximation becomes better as the invariant mass of colliding baryons increases. High energy pions may thus be a probe of the equation of state less impacted by this type of model uncertainties. 

Inelastic nucleon-nucleon cross-sections are modified in-medium by using a scaling factor that depends on the effective masses of the scattering baryons, in agreement with the results
of the one-pion exchange microscopical model of Ref. \cite{Larionov:2003av}. Within this model in-medium modified inelastic $NN\rightarrow N\Delta$ cross-sections have been determined by including effects such as in-medium corrections to the pion propagator, vertex corrections and in-medium effective masses. The dominant effect could be described by a correction factor depending on effective masses of initial and final state-baryons and of the medium-modified invariant mass obtained by replacing canonical with kinetic momenta.

The dynamics of the present model is non-relativistic and consequently modifications of the invariant mass using a relativistic mean field approach is not possible. Instead we follow the approach in Ref.~\cite{Cozma:2014yna} developed to ensure total energy conservation of the system, which naturally leads to threshold effects and in-medium modifications of cross-sections. The central assumption of the approach is that no true two-body scattering processes exist, but rather they are modified by interaction with the rest of the system. Due to energy exchange with the fireball the initial- and final-state invariant masses of the two scattering particles ($s_{ini}$ and $s_{fin}$), determined using vacuum masses and momenta, differ. Considering the fact that vacuum inelastic cross-section for resonance excitation increases with the invariant mass, the contribution involving two-particles scattering with the higher invariant mass dominates the total scattering amplitude. This approximation is best close to threshold and was estimated to be valid up to impact energies of about 800 MeV/nucleon. Therefore the medium modified invariant mass used to determine cross-sections reads $ s^*=Max(s_{ini},s_{fin})$. In Ref.~\cite{Cozma:2014yna} it was shown that $s_{fin}-s_{ini}>0$ for EoS'es that are not too soft ($L>$0 MeV). The above Ansatz thus translates into contributions that involve energy exchanges with the fireball in the initial state, followed by inelastic scattering of the two baryons, dominating the total scattering amplitudes of resonance excitation.

The expression for the in-medium inelastic cross-sections thus reads
\begin{eqnarray}
\sigma_{NN\to N\Delta}^{(med)}(s^*)=\frac{\mu^{(ini)*}}{\mu^{(ini)}}\,\frac{\mu^{(fin)*}}{\mu^{(fin)}}\,\sigma_{NN\to N\Delta}^{(vac)}(s^*)
\end{eqnarray}
with starred and regular variables corresponding to in-medium and vacuum quantities and $\mu$ denoting the reduced mass of the system. A similar expression for the modification factor was obtained in Refs. \cite{Schulze:1997zz,Persram:2001dg,Li:2005jy} on qualitative grounds for elastic nucleon-nucleon cross-sections. For effective masses the non-relativistic formula is used e.g. $m^*={m}/{(1.0+\frac{m}{p}\frac{dU}{dp})}$. The density dependence of effective masses has only a rather small impact on pion multiplicities, in spite of modification factors that amount to values in the range of 0.5-0.7 at saturation. Such substantial decreases of cross-sections are partially compensated by having also smaller absorption $N\Delta\to NN$ rates. The impact of in-medium modifications of inelastic cross-sections on pion observables due to isospin asymmetry dependence of effective masses were found to be small during tests and have been therefore neglected in the present study.

The cross-section for the resonance absorption reaction $NR\rightarrow NN$ is determined using a detailed balance formula \cite{Danielewicz:1991dh},
\begin{eqnarray}
 &&\frac{d\sigma^{(NR\rightarrow NN)}}{d\Omega}(s^*)=\frac{1}{4}\frac{m_R\,p_{NN}^2}{p_{NR}}\frac{d\sigma^{(NN\rightarrow NR)}}{d\Omega}(s^*)\times \\
  &&\qquad\quad\bigg(\frac{1}{2\pi}\int_{m_N+m_\pi}^{\sqrt{s_{ini}}-m_N}dM M\,p'_{NR}\,A_R(M)\bigg)^{-1}. \nonumber
\end{eqnarray}
Due to the difference between $s_{ini}$ and $s_{fin}$ momenta $p_{NN}$ and $p_{NR}$ have to be evaluated using the invariant masses of the $NN$ (final) and $NR$(initial) states respectively. Such a prescription can be understood since, in the expression for the cross-section of a 2-body reaction $NR\rightarrow NN$, $p_{NR}$ originates from the evaluation of the incoming flux, while $p_{NN}$ arises from the final-state phase space.

The pion decay width of resonances is determined using the expression \cite{Weil:2016zrk}
\begin{eqnarray}
 \Gamma_{R\rightarrow N\pi}(\sqrt{s})=\Gamma_{R\rightarrow N\pi}(\sqrt{s_0})\,\frac{\sqrt{s_0}}{\sqrt{s}}\,
 \frac{p^3}{p_0^3}\,\frac{p_0^2+\Lambda^2}{p^2+\Lambda^2},
\end{eqnarray}
depending on the invariant mass $\sqrt{s}$ and its pole mass value $\sqrt{s_0}$; $p$ and $p_0$ are the corresponding pion
momenta in the rest frame of the resonance. The above formula is a particular case of a more general
expression~\cite{Manley:1992yb} for a value
of the orbital angular momentum of the pion-nucleon system equal to 1.
The quantity $\Lambda$ is computed using
\begin{eqnarray}
\Lambda=\sqrt{(m_R-m_{N}-m_{\pi})^2+\Gamma^2/4.0},
\end{eqnarray}
where $m_R=1.232$ GeV and $\Gamma=0.115$ GeV (pole mass properties of the resonance, $\Delta(1232)$ in this case);
similarly for other resonances (N(1440), etc). In the parent TuQMD model, as well a in previous publications \cite{Cozma:2016qej,Cozma:2014yna}, a formula for the width that is close to the Huber parametrization \cite{Huber:1994ee} had been used. It leads to pion absorption cross-sections close to threshold that are too large, by a factor close to 2, as compared to the experimental data. At invariant masses in the vicinity of the resonance's mass pole realistic values are obtained. The above parametrization for the resonance decay width solves the mentioned problem. It is worth noting that a modification of the resonance decay width does not require a refit of the Huber OBE model as long as double $\Delta$ production is negligible, since the difference can be absorbed in the $\pi N\Delta$ vertex form-factor.

The above expression for the decay width employs a generic variable $s$. For the resonance decay $R\rightarrow N\pi$ and pion absorption $\pi N\rightarrow R$ terms in the transport model the expression is evaluated using a modified invariant mass $s^*=Max(s_{ini},s_{fin})$ supplemented by the same argumentation as for baryon-baryon scattering.

Contributions of pion optical potentials have been included by using the Ericson-Ericson parametrization to describe their density, isospin asymmetry and momentum dependence, see Ref. \cite{Cozma:2016qej} for all relevant details. The set of parameter values for the optical potential commonly known as Batty-1 \cite{Batty:1978aa} has been used extensively in this work, with one exception. In \secref{4} the effective S-wave model set of parameters (denoted S') \cite{Cozma:2016qej} has been used to study the residual model dependence on $p_T$ spectra of PMR. Mean field propagation of pions is treated similarly to that of nucleons, by associating a Gaussian wave function to them, whose width has been set such that the ratio of pion-to-proton charge radii is close to its experimental value \cite{Cozma:2016qej}.

Threshold effects have been accounted for within the global energy conservation (GEC) scenario introduced in Ref. \cite{Cozma:2014yna} and which has been briefly presented above. It has been checked that such a scenario is compatible with a system of nucleons, $\Delta$(1232)s and pions reaching chemical equilibrium. Specifically, this has been achieved by performing numerical checks of detailed balance. To this end, nuclear matter in a box at temperature T=60 MeV has been simulated using the full model. The initial state of the system consisted of nucleons and pions with relative multiplicity abundances of 90$\%$ and 10$\%$ respectively. Detailed balanced for the reactions $N+N\leftrightarrow N+\Delta$ and $\Delta \leftrightarrow \pi N$ was shown to be fulfilled at a few percent level after a time lapse of about 100 fm/c which signals that chemical equilibrium has been reached. With appropriate settings the model reproduces the benchmark results of the TMEP Collaboration \cite{Xu:2016lue,Zhang:2017esm,Ono:2019ndq}.

%----------------------------------------------
\subsection{Baryon in-medium interactions}
\seclab{2b}
The same parametrization for the equation of state of nuclear matter as in ~\cite{Cozma:2017bre} is used.
The potential part reads
\begin{eqnarray}
\eqlab{eos}
\frac{E}{N}(\rho,\beta)&=&A_u\frac{\rho(1-\beta^2)}{4\rho_0}+A_l\frac{\rho(1+\beta^2)}{4\rho_0} \\
&&+\frac{B}{\sigma+1}\frac{\rho^{\sigma}}{\rho_0^\sigma}\,(1-x\beta^2)+
\frac{D}{3}\frac{\rho^2}{\rho_0^2}\,(1-y\beta^2) \nonumber\\
&&+\frac{1}{\rho\rho_0}\sum_{\tau,\tau'} C_{\tau \tau'}\!\!\int\!\!\int d^{\!\:3} \vec{p}\,d^{\!\:3} \vec{p}\!\;'
\frac{f_\tau(\vec{r},\vec{p}) f_{\tau'}(\vec{r},\vec{p}\!\;')}{1+(\vec{p}-\vec{p}\!\;')^2/\Lambda^2}. \nonumber
\end{eqnarray}
Its analytic form is similar to MDI Gogny-inspired parametrizations~\cite{Das:2002fr,Xu:2014cwa}, but differs from
these by an extra density-dependent but momentum-independent term, proportional to the $D$ parameter, that has been introduced
in order to allow independent variations of the slope $L$ and curvature $K_{sym}$ parameters of the symmetry energy, while
keeping the neutron-proton isovector effective mass difference fixed. 

\begin{figure*}[htb]
 \includegraphics[width=0.495\textwidth]{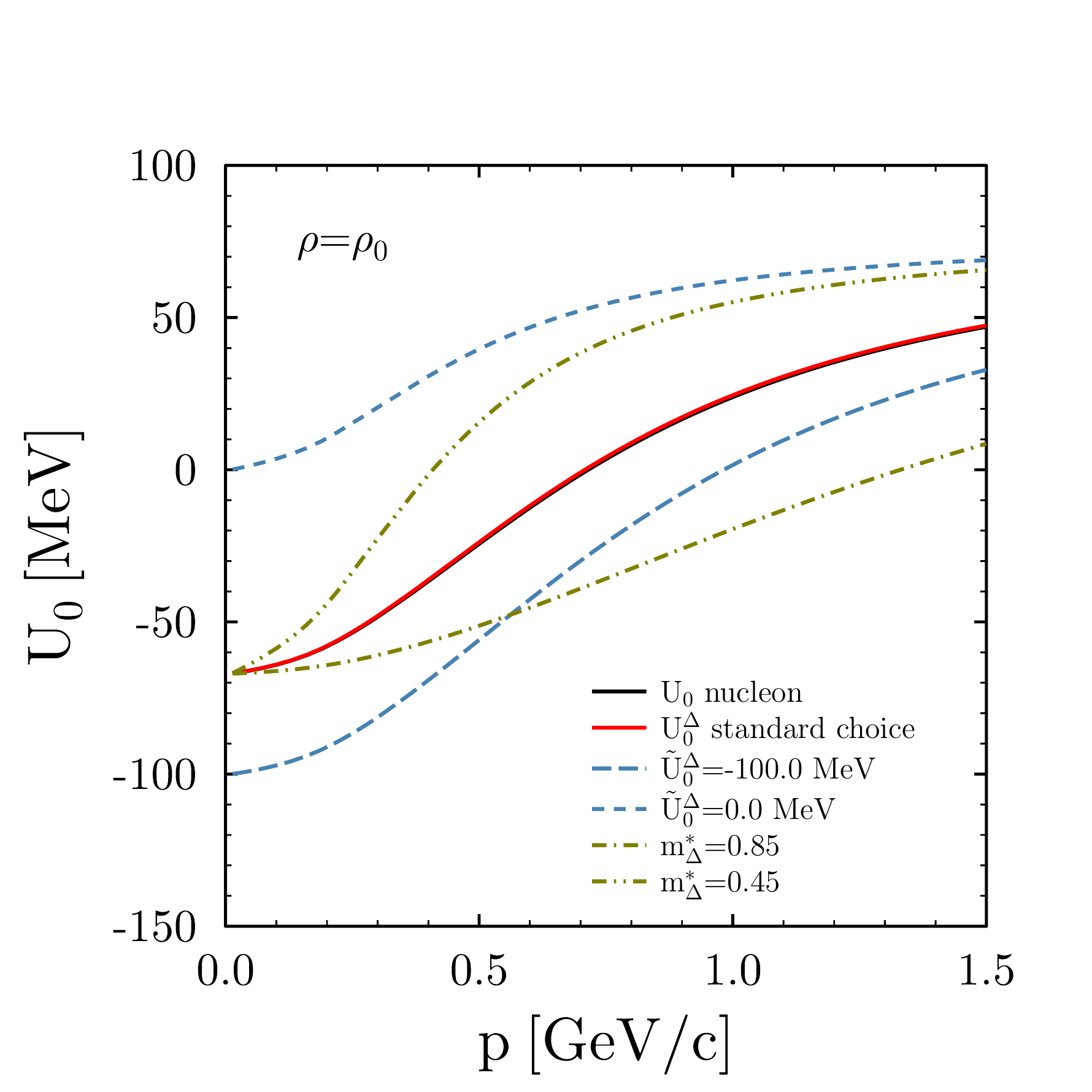}	
 \includegraphics[width=0.495\textwidth]{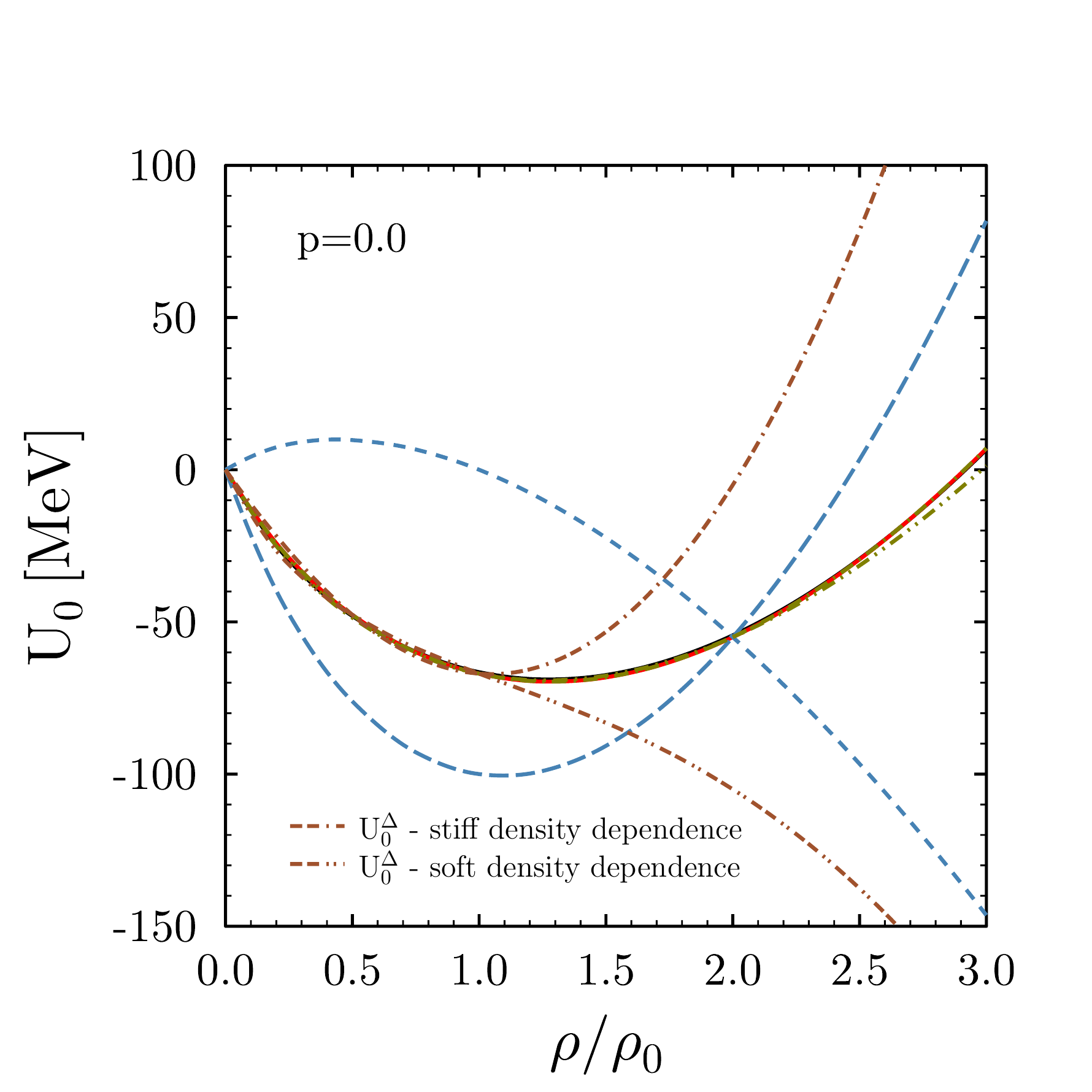}
 \caption{Momentum (left panel) and density dependence (right panel) of ISDP for several choices of depth and effective isoscalar mass, as discussed in the text, compared to the nucleon isoscalar potential with the compressibility modulus set to $K_0$=245 MeV. Each explanatory key applies to both plots. In the left panel, the nucleon potential in symmetric matter $U_0$ and the standard choice $U_{0}^\Delta$ almost coincide. In the right, panel the ISDPs $U_{0}^\Delta$ corresponding to different effective masses show similar density dependence.}
 \figlab{iscpot}
\end{figure*}

The corresponding single-particle nucleon potential is given by

\begin{eqnarray}
\eqlab{sympot}
 U_\tau(\rho,\beta,p)&=&A_u\frac{\rho_{\tau'}}{\rho_0}+A_l\frac{\rho_{\tau}}{\rho_0} \\
&&+B\,\Big(\frac{\rho}{\rho_0}\Big)^\sigma(1-x\beta^2)
+8\tau x\frac{B}{\sigma+1}\frac{\rho^{\sigma-1}}{\rho_0^\sigma}\beta\rho_{\tau'} \nonumber\\
&&+D\,\Big(\frac{\rho}{\rho_0}\Big)^2(1-y\beta^2)
+8\tau y\frac{D}{3}\frac{\rho}{\rho_0^2}\beta\rho_{\tau'} \nonumber\\
&&+\frac{2C_{\tau \tau}}{\rho_0}\int d^{\!\:3} \vec{p}\!\;'\, \frac{f_\tau(\vec{r},\vec{p}\!\;')}{1+(\vec{p}-\vec{p}\!\;')^2/\Lambda^2} \nonumber\\
&&+\frac{2C_{\tau \tau'}}{\rho_0}\int d^{\!\:3} \vec{p}\!\;'\, \frac{f_{\tau'}(\vec{r},\vec{p}\!\;')}{1+(\vec{p}-\vec{p}\!\;')^2/\Lambda^2}. \nonumber
\end{eqnarray}
In the above expressions $\rho$, $\beta$ and $p$ denote the density, isospin asymmetry and momentum variables respectively.
The label $\tau$ designates the isospin component of the nucleon and takes the value $\tau$=-1/2 (1/2) 
for neutrons (protons). For cold nuclear matter it holds $f_\tau(\vec{r},\vec{p})=(2/h^3)\Theta(p_F^\tau-p)$, with $p_F^\tau$ the 
Fermi momentum of nucleons with isospin $\tau$.

\begin{figure*}[htb]
 \includegraphics[width=0.495\textwidth]{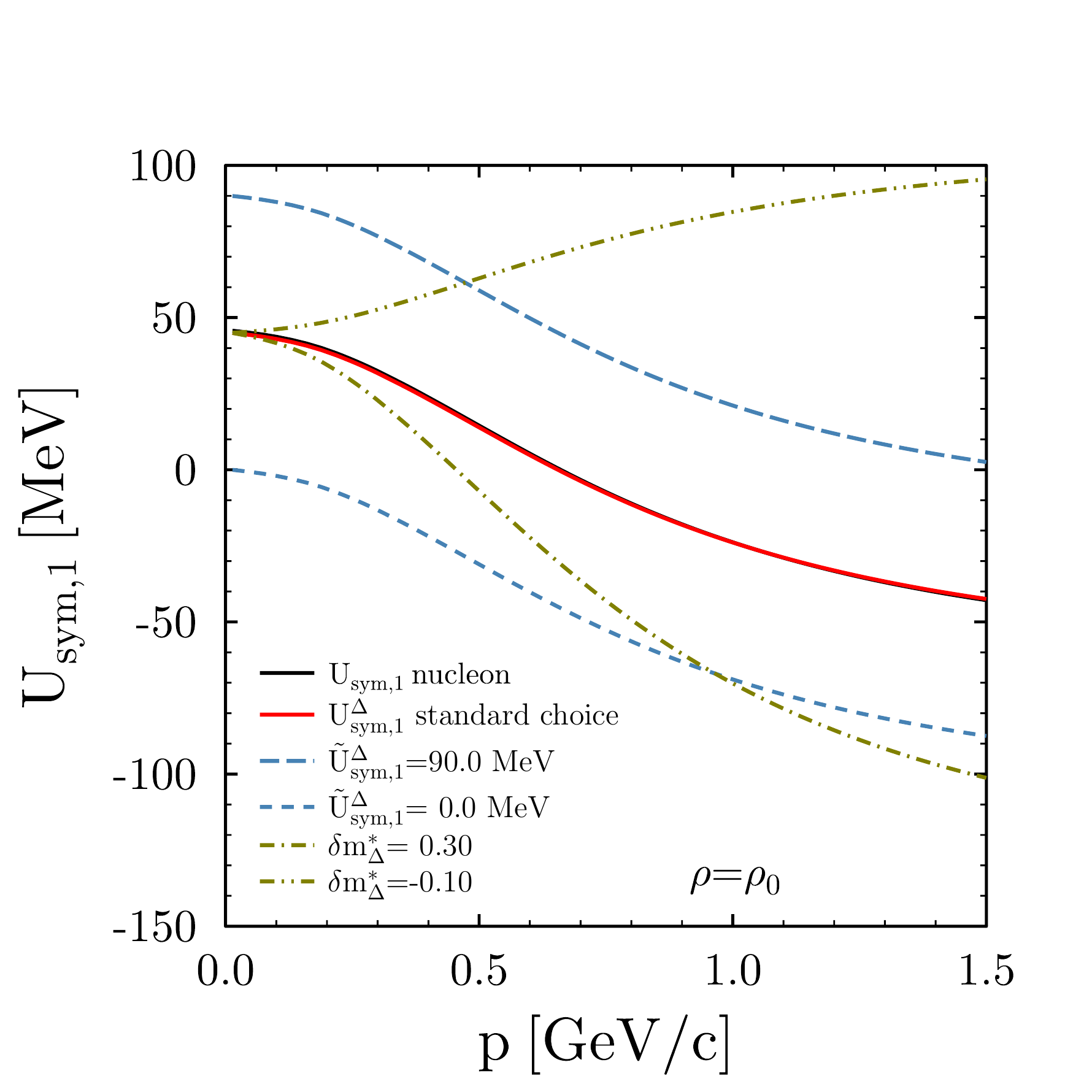}	
 \includegraphics[width=0.495\textwidth]{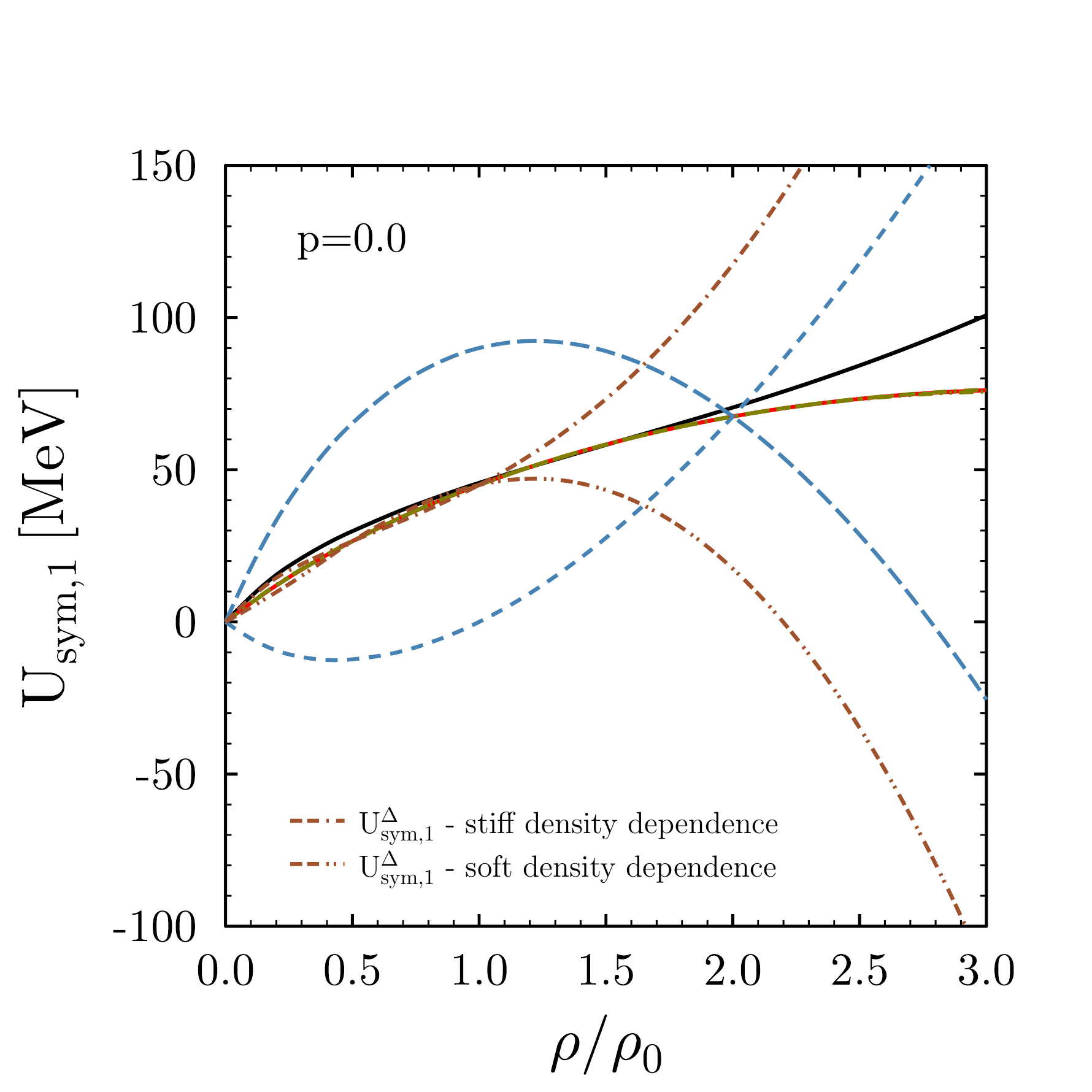}
 \caption{Momentum (left panel) and density dependence (right panel) of the leading order symmetry potential $U^{\Delta}_{sym,1}$.  The leading order nucleon symmetry potential corresponding to $L$=60.5 MeV and $K_{sym}$=-81.0 MeV is also shown for comparison. Each explanatory key applies to both plots. In the left panel, the leading order nucleon symmetry potential $U_{sym,1}$ and the standard choice IVDP $U^{\Delta}_{sym,1}$ almost coincide. In the right, panel the IVDPs $U^{\Delta}_{sym,1}$ corresponding to different effective mass differences show a very similar density dependence. }
 \figlab{isvpot}
\end{figure*}

It is common practice, within the framework of transport models, to set the resonance potentials in terms of the
nucleonic one. This choice is guided by the decay channels of the resonance in question into a final state
comprising a nucleon and a pion \cite{Li:2002yda}. This approach is particularly well suited for the $\Delta$(1232) baryon
which has a branching ratio close to 1 for the $\Delta\rightarrow N\pi$ decay channel.
It is nevertheless applied to the entire list of resonances included in the given transport model. To be specific,
\begin{eqnarray}
U^R_\tau(\rho,\beta,p)&=&\frac{1}{2}(1-\tau/T)\;U_{-\frac{1}{2}}(\rho,\beta,p)\\&+&\frac{1}{2}(1+\tau/T)\;U_{\frac{1}{2}}(\rho,\beta,p),\nonumber
\eqlab{respot}
\end{eqnarray}
where $T$ and $\tau$ are the isospin and its desired projection for the resonance in question; $U_{-\frac{1}{2}}$ and
$U_{\frac{1}{2}}$ represent the neutron and proton potentials respectively, whose expressions can be read from \eqref{sympot}.
For an isospin $T$=3/2 resonance it leads to
\begin{eqnarray}
 \begin{array}{lcrcrcrcrl}
U_{\Delta^-}&=& U_{-\frac{1}{2}}&& &=& U_{is}&+&U_{iv}&\\
U_{\Delta^0}&=&\frac{2}{3}\,U_{-\frac{1}{2}}&+&\frac{1}{3}\,U_{\frac{1}{2}}&=& U_{is}&+&\frac{1}{3}U_{iv}&\\
U_{\Delta^+}&=&\frac{1}{3}\,U_{-\frac{1}{2}}&+&\frac{2}{3}\,U_{\frac{1}{2}}&=& U_{is}&-&\frac{1}{3}U_{iv}&\\
U_{\Delta^{++}}&=&&&U_{\frac{1}{2}}&=& U_{is}&-&U_{iv}&,
\eqlab{deltapot}
\end{array}
\end{eqnarray}
%\begin{eqnarray}
% \begin{array}{lcrcr}
%U_{\Delta^-}&=& U_{-\frac{1}{2}}&&\\
%U_{\Delta^0}&=&\frac{2}{3}\,U_{-\frac{1}{2}}&+&\frac{1}{3}\,U_{\frac{1}{2}}\\
%U_{\Delta^+}&=&\frac{1}{3}\,U_{-\frac{1}{2}}&+&\frac{2}{3}\,U_{\frac{1}{2}}\\
%U_{\Delta^{++}}&=&&&U_{\frac{1}{2}}
%\eqlab{deltapot}
%\end{array}
%\end{eqnarray}
%
%\begin{eqnarray}
%\begin{array}{lcrcr}
% U_{\Delta^-}&=& U_{is}&+&U_{iv} \\
% U_{\Delta^0}&=& U_{is}&+&\frac{1}{3}U_{iv}\\
% U_{\Delta^+}&=& U_{is}&-&\frac{1}{3}U_{iv}\\
% U_{\Delta^{++}}&=& U_{is}&-&U_{iv}
% \end{array}
%\end{eqnarray}
which can be split into iso-scalar and iso-vector contributions, denoted
above by $U_{is}$ and $U_{iv}$. Their expression can be readily found out to be
\begin{eqnarray}
\eqlab{iscisvpot1}
 U_{is}(\rho,\beta,p)&=&\frac{A_u+A_l}{2}\,\frac{\rho}{\rho_0}+B\,\Big(\frac{\rho}{\rho_0}\Big)^\sigma\,(1-x\beta^2)\\
 &&+D\,\Big(\frac{\rho}{\rho_0}\Big)^2\,(1-y\beta^2)\nonumber\\
 &&+\frac{C_l+C_u}{\rho_0}\,[\;I(p,p_F^n)+I(p,p_F^p)\,],\nonumber\\
 \eqlab{iscisvpot2}
 U_{iv}(\rho,\beta,p)&=&\frac{A_l-A_u}{2}\,\frac{\rho}{\rho_0}\,\beta
 -2x\frac{B}{\sigma+1}\Big(\frac{\rho}{\rho_0}\Big)^\sigma\beta \\
 &&-2y\frac{D}{3}\Big(\frac{\rho}{\rho_0}\Big)^2\beta\nonumber\\
 &&+\frac{C_l-C_u}{\rho_0}\,[\;I(p,p_F^n)-I(p,p_F^p)\,],\nonumber
\end{eqnarray}
with the following notations: $C_l=C_{1/2,1/2}=C_{-1/2,-1/2}$, $C_u=C_{1/2,-1/2}=C_{-1/2,1/2}$, $p_F^n$ and $p_F^p$
represent the Fermi momenta of neutrons and protons respectively; $I(p,p_F^\tau)$ stands
for the integrals appearing in \eqref{sympot}, for which an analytic expression can be derived for the case of zero-temperature nuclear matter
\begin{eqnarray}
&&I(p,p_F^\tau)=\int d^{\!\:3} \vec{p}\!\;'\, \frac{f_\tau(\vec{r},\vec{p}\!\;')}{1+(\vec{p}-\vec{p}\!\;')^2/\Lambda^2}\\
&&=\frac{2\pi}{h^3}\Lambda^3\Bigg [ \frac{\Lambda^2+p_F^2(\tau)-p^2}{2\Lambda p}\,\mathrm{ln}\,
 \frac{\Lambda^2+[p+p_F(\tau)]^2}{\Lambda^2+[p-p_F(\tau)]^2} \nonumber\\
&&+\frac{2p_F(\tau)}{\Lambda}+ 2\,\Bigg ( \mathrm{arctan} \frac {p-p_F(\tau)}{\Lambda} 
-\mathrm{arctan} \frac {p+p_F(\tau)}{\Lambda} \Bigg ) \Bigg ]. \nonumber
\end{eqnarray}
It can be easily seen that the expression above is an odd function of $p_F^\tau$. As a result the isoscalar and
isovector potentials above are even and odd functions in the isospin asymmetry variable $\beta$ respectively, as required by
charge symmetry. 

It is worth stressing that in the above equations $U_{is}$ and $U_{iv}$ are identical to the corresponding nucleonic potentials, as a direct consequence of the Ansatz in \eqref{respot}, and the parameters appearing in their expressions are therefore determined by reproducing nuclear matter properties.

The nucleonic potential in \eqref{sympot} can be expanded in a Taylor series in terms of the isospin asymmetry parameter around the point $\beta$=0
\begin{eqnarray}
U_\tau(\rho,\beta,p)&=&U_0(\rho,p)+\sum_{i=1,\infty}\,U_{sym,i}(\rho,p)\,(2\tau\,\beta)^i\,. 
\eqlab{taylorexppot}
\end{eqnarray}
The first two terms, $U_0(\rho,p)$ and $U_{sym,1}(\rho,p)$, represent the nucleon potential in isospin symmetric nuclear matter
and the first-order symmetry potential respectively. Their expressions can be derived from those for the isoscalar and isovector nucleon potentials in \eqref{iscisvpot1} and \eqref{iscisvpot2} using the relations
\begin{eqnarray}
U_0(\rho,p) &=& U_{is}(\rho,\beta=0,p)\,,  \\
U_{sym,1}(\rho,p) &=& \lim_{\beta\to 0} \frac{U_{iv}(\rho,\beta,p)}{\beta}\,. \nonumber
\end{eqnarray}
Naturally, for the case of the Ansatz used in \eqref{respot} we have $U_0^R(\rho,p)=U_0(\rho,p)$ while $U_{sym,1}^R(\rho,p)$=$U_{sym,1}(\rho,p)$ once the replacement $2\tau \rightarrow \tau/T$ is made in \eqref{taylorexppot}.

In this study we depart from the usually made assumption in transport models, briefly presented above, that $U_{is}$ and $U_{iv}$ entering \eqref{deltapot} are the corresponding nucleon potentials. We do however assume that their expressions in terms of density, isospin asymmetry and momentum are the same but different values for the coupling parameters. In order to make this distinction clear we add a superscript ``$\Delta$'' to relevant quantities, in particular $U_{is}^\Delta$, $U_{iv}^{\Delta}$, $U_0^\Delta$ and $U_{sym,1}^\Delta$. We allow the freedom that the density and momentum dependence of resonance potentials be different at intermediate and long ranges as well as at densities below twice saturation density. We do however require, for a standard case labeled accordingly where distinction is relevant, that their high density part is similar to that of nucleons, in view of their similar quark structure. This approach is different from the one pursued in Refs.~\cite{Cozma:2016qej,Cozma:2014yna} where both the isoscalar and isovector components of the DPOT were modified by a scaling factor. 

In the following we present details of how the values of parameters entering in \eqref{iscisvpot1} and \eqref{iscisvpot2} are fixed in this study. There are six free parameters entering the expression of the ISDP $U_{is}^\Delta$: $(A_l+A_u)/2$, $B$,
$\sigma$, $D$, $C_u+C_l$ and $\Lambda$ (a superscript ``$\Delta$'' is in order for each of these parameters, but is omitted). For simplicity we set $D$=0.0 MeV and $\sigma$=1.465. The remaining four are determined by requiring that certain values for the isoscalar effective mass of the resonance $m_\Delta^*$ and the potential in symmetric matter at suitable values for density and momentum, $\tilde{U}_0^\Delta\equiv U_0^\Delta(\rho_0,p=0)$, $U_0^\Delta(2\rho_0,p=0)$ and $U_0^\Delta(\rho_0,p=\infty)$, are described. The quoted value for $\sigma$ ensures that the density dependence of the resulting ISDP is close to that of the nucleon once the
values at the three above mentioned points fulfill this requirement too.

The expression of the IVDP $U_{iv}^\Delta$ contains four additionally free parameters: $(A_l-A_u)/2$, $C_l-C_u$, $x$
and $y$ (again, the ``$\Delta$'' superscript is omitted). The value of the last one is irrelevant in the context of setting $D$=0 MeV. The remaining three are determined by requiring definite values for $\tilde{U}_{sym,1}^\Delta\equiv U_{sym,1}^\Delta(\rho_0,p=0)$, $U_{sym,1}^\Delta(2\rho_0,p=0)$ and the isovector mass-splitting $\delta m^*_{\Delta}=(m^*_{\Delta^{-}}-m^*_{\Delta^{++}})/m_\Delta$, the last quantity being evaluated at saturation density and $\beta$=0.5. The second order symmetry potential $U_{sym,2}^\Delta$ impacts the value of the isovector mass-splitting at a few percent level since its contribution to the symmetry potential is smaller than 10$\%$ irrespective of the value of $\beta$.

The values for the ten model parameters for the case when the DPOT is similar to the
nucleon's up to twice saturation density and for kinetic energies up to 1.0 GeV are presented in \tabref{model_input_params}.
For the isovector part, the quoted parameter values lead to nucleon in-medium interactions that correspond to a density dependence of SE with a slope $L$=60.5 MeV and curvature parameter $K_{sym}$=-81.0 MeV.
%This is close to the world average for the stiffness of thesymmetry energy \textcolor{red}{[citation needed]}.

\begin{table}
\caption{Input quantities and their values (first and second columns) used to set the DPOT
together with the model parameters appearing in~\eqref{iscisvpot1} and \eqref{iscisvpot2} and their determined values
(third and fourth columns). This set of parameters leads to ISDP and IVDP that resemble the nucleonic potentials closely. Quantities denoted by capital letters are expressed in units of MeV, while the rest are dimensionless. The effective mass $m_\Delta^*$ is expressed in units relative to the vacuum value of the mass of the $\Delta$(1232) isobar.}
\tablab{model_input_params}       % Give a unique label
\begin{center}
\begin{tabular}{cr@{.}l|cr@{.}l}
\hline\noalign{\smallskip}
\multicolumn{3}{c|}{Input} & \multicolumn{3}{c}{Parameters}\\
\noalign{\smallskip}\hline\noalign{\smallskip}
$m_{\Delta}^*$ & 0&65 & $\Lambda$ & 700&98\\
$U_0^\Delta(\rho_0,p=0)$ & -67&0 & $C_l+C_u$ & -153&82\\
$U_0^\Delta(2\rho_0,p=0)$ & -55&0 & $A_l+A_u$ & -26&15 \\
$U_0^\Delta(\rho_0,p=\infty)$ & +75&0 & $B$ & 88&08 \\
 & & & $D$ (fixed) & 0&0 \\
 & & & $\sigma$ (fixed) & 1&465 \\
\noalign{\smallskip}\hline\noalign{\smallskip}
$\delta m_{\Delta}^*$ & 0&175 & $C_l-C_u$ & 125&50 \\
$U_{1,sym}^\Delta(\rho_0,p=0)$ & +45&0 & $A_l-A_u$ &  -109&97 \\
$U_{1,sym}^\Delta(2\rho_0,p=0)$ & +67&5 & $x$ & 0&140 \\
 & & & $y$    (fixed) & 0&0 \\
\noalign{\smallskip}\hline
\end{tabular}
\end{center}
%\vspace*{2cm}  % with the correct table height
\end{table}

In \figref{iscpot} the momentum and density dependence of ISDP in symmetric nuclear matter $U_0^{\Delta}$
for several cases is presented. The corresponding nucleon potential, $U_0$, is also shown for reference. A standard
$U_0^{\Delta}$ that corresponds to a potential depth at saturation and zero momentum 
$\tilde{U}_0^{\Delta}$=-67.0 MeV and an isoscalar effective mass $m^*_\Delta$ = 0.65 has been defined.
It mirrors both the momentum and density dependence of the nucleon $U_0$ potential, as can be seen from the left and
right panels of \figref{iscpot} respectively. Sensitivity of pionic observables to $U_0^{\Delta}$ will be studied by varying
its depth at saturation $\tilde U_0^{\Delta}$ in the interval [-100.0,0.0] MeV and the isoscalar effective mass in the range [0.45,0.85]. The potentials corresponding to the limits of these intervals are shown in \figref{iscpot}. Modification of the ISDP depth at saturation induces also a drastic change of the density dependence. Additionally, two potentials denoted
as ``stiff density dependence'' and ``soft density dependence'' are also shown. 
They have been constructed by modifying the value of the potential at twice saturation density $U^\Delta_{0}(2\rho_0,p=0)$ 
to -5 MeV for stiff and -105 MeV for soft and allowing for
a non-zero value of the $D$ parameter while keeping parameters $\sigma$ and $y$ fixed to the values quoted
in \tabref{model_input_params}. This procedure ensures that the IVDP remains unchanged.
The model parameters have been adjusted such as to modify only the density dependence above saturation, while keeping the potential depth at saturation and half-saturation (both at zero momentum) and the isoscalar effective mass fixed. These two potentials will be used to study the impact of stiff and soft supranormal density dependence of the $U_0^\Delta$ potential on pionic observables.

Similarly, in \figref{isvpot} the momentum and density dependence of the leading order symmetry potential of $\Delta$(1232) $U^\Delta_{sym,1}$ is shown. A standard choice $U^\Delta_{sym,1}$ potential is defined by requiring that its strength at saturation and zero momentum is $\tilde{U}^\Delta_{sym,1}$=45.0 MeV and the isovector mass splitting amounts to $\delta m^*_{\Delta}$=0.175. The corresponding nucleon potential, that leads to a density dependence of symmetry energy with a slope $L$=60.5 MeV and curvature parameter $K_{sym}$=-81.0 MeV, is shown for comparison. Sensitivity of pionic observables
to $U_{sym,1}^{\Delta}$ will be studied by varying its strength at saturation $\tilde{U}^\Delta_{sym,1}$ in the interval [-15.0,90.0] MeV and the isovector mass splitting $\delta m^*_{\Delta}$ in the range [-0.10,0.30]. Also in this case, two potentials labeled ``stiff'' and
``soft'' density dependence have been constructed by modifying the potential strength at twice saturation density, while keeping the values at saturation and half-saturation fixed, all this at p=0. The choices of $U^\Delta_{sym,1}(2\rho_0,p=0)$ equal to 117.5 MeV and 17.5 MeV have been made for the stiff and soft cases respectively. Technically this was achieved by modifying the value of quantity $D\,y$ (redefined as a variable independent of $D$) while keeping $D$ equal to zero. 

\begin{figure*}[htb]
 \includegraphics[width=0.495\textwidth]{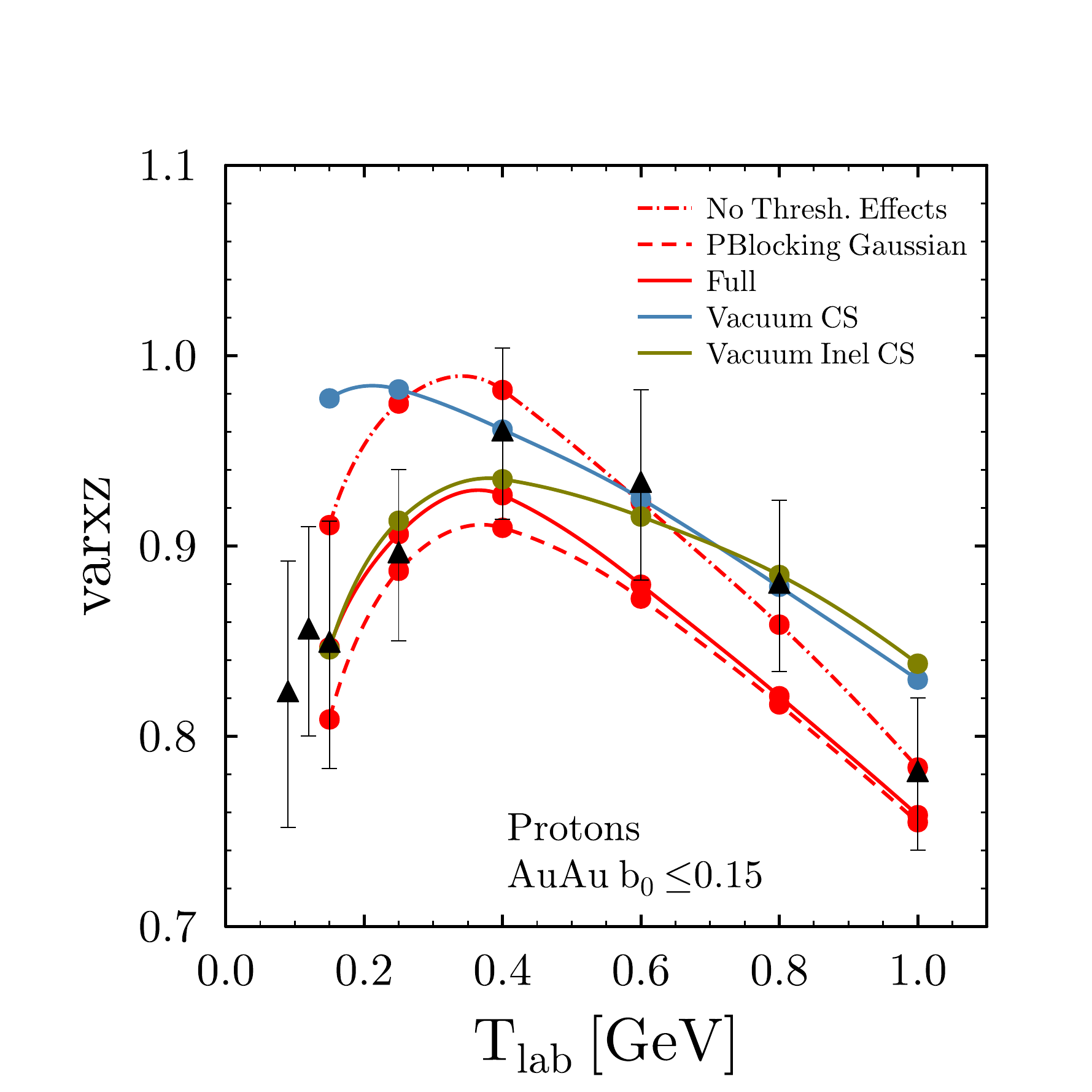}	
 \includegraphics[width=0.495\textwidth]{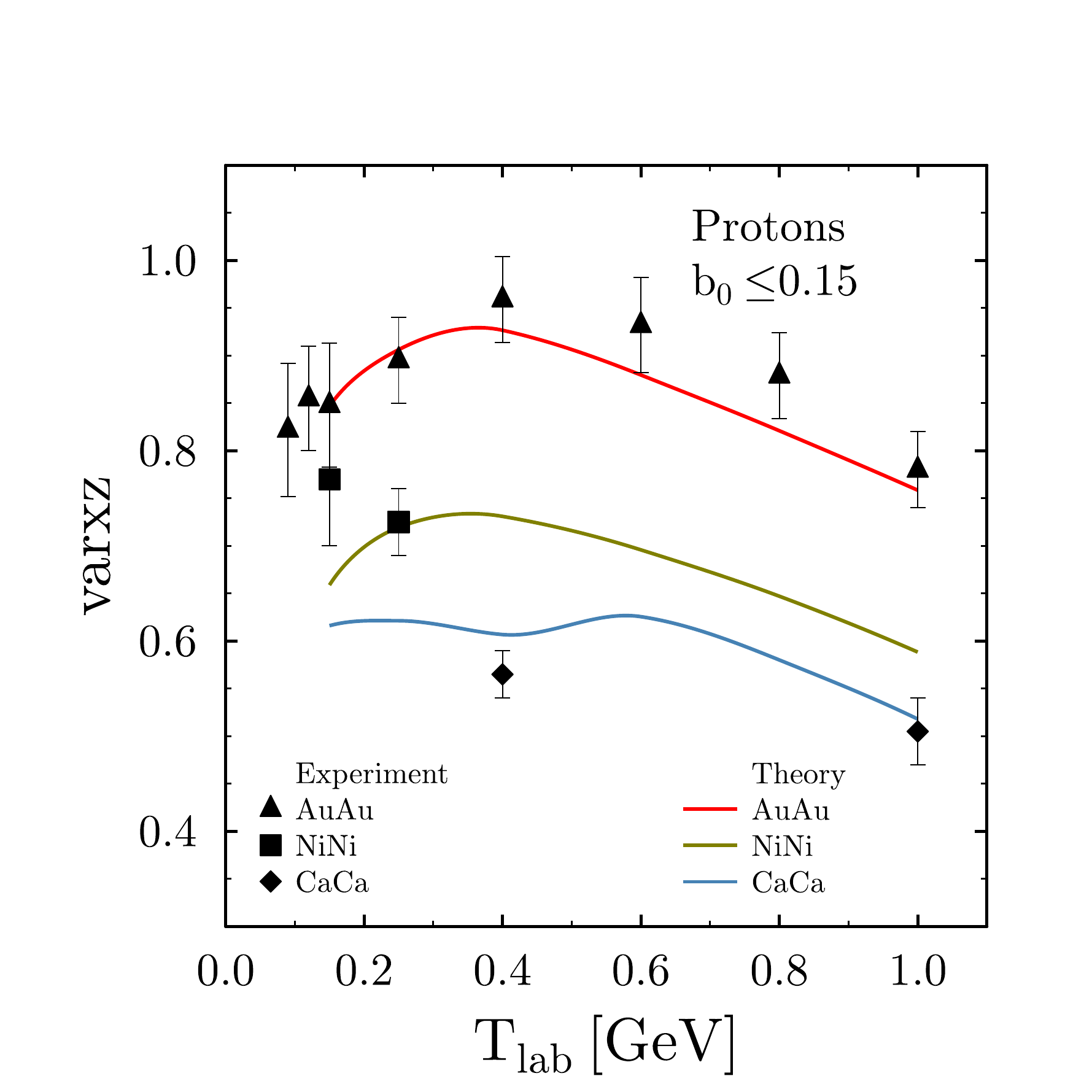}
 \caption{(Left Panel) Model dependence for proton stopping in central AuAu collisions as a function of the impact energy per nucleon. (Right Panel) System size dependence of stopping for protons in central collision for systems of different masses. The FOPI experimental data~\cite{Reisdorf:2010aa} have been plotted for  comparison.}
 \figlab{stopping}
\end{figure*}

\begin{figure*}[htb]
 \includegraphics[width=0.495\textwidth]{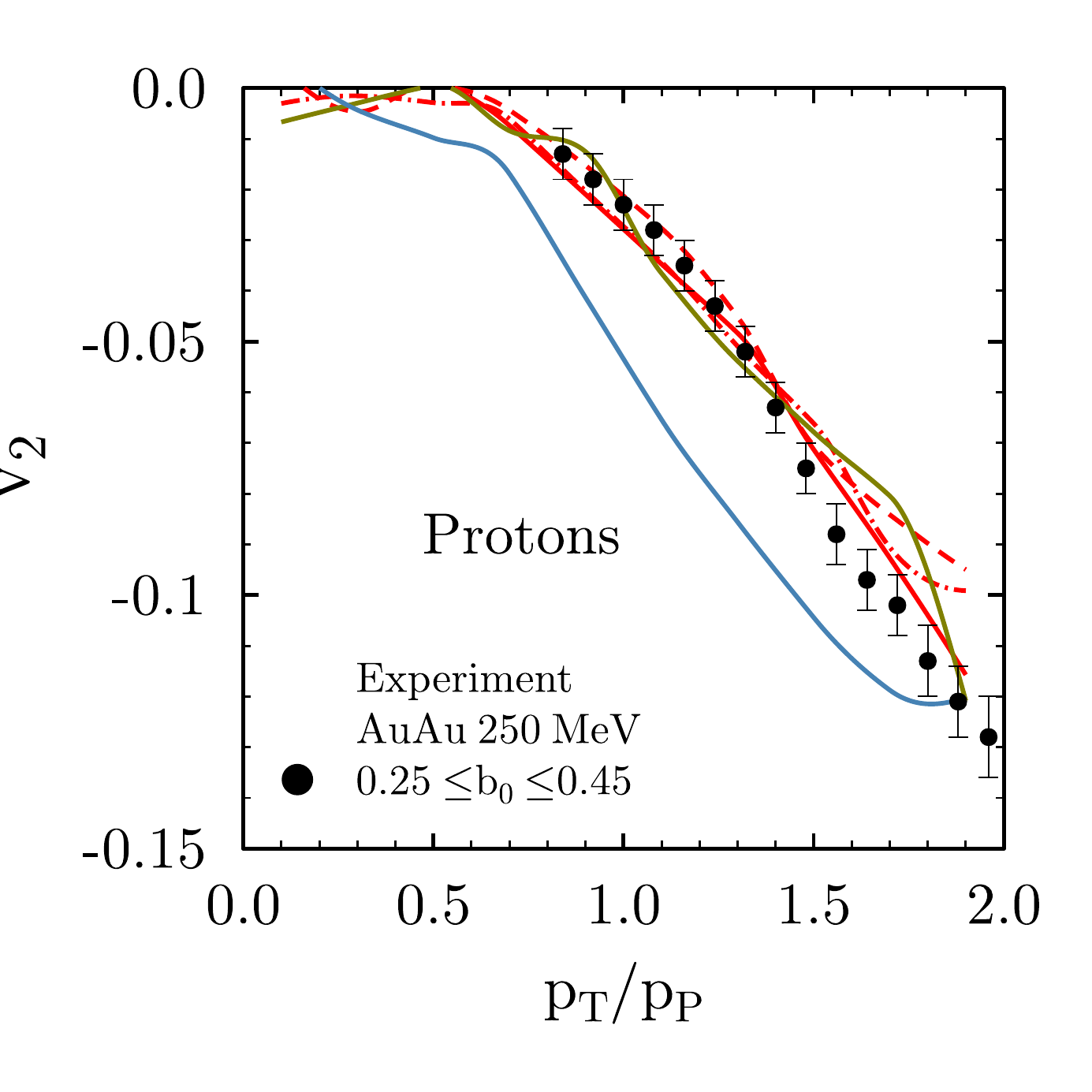}	
 \includegraphics[width=0.495\textwidth]{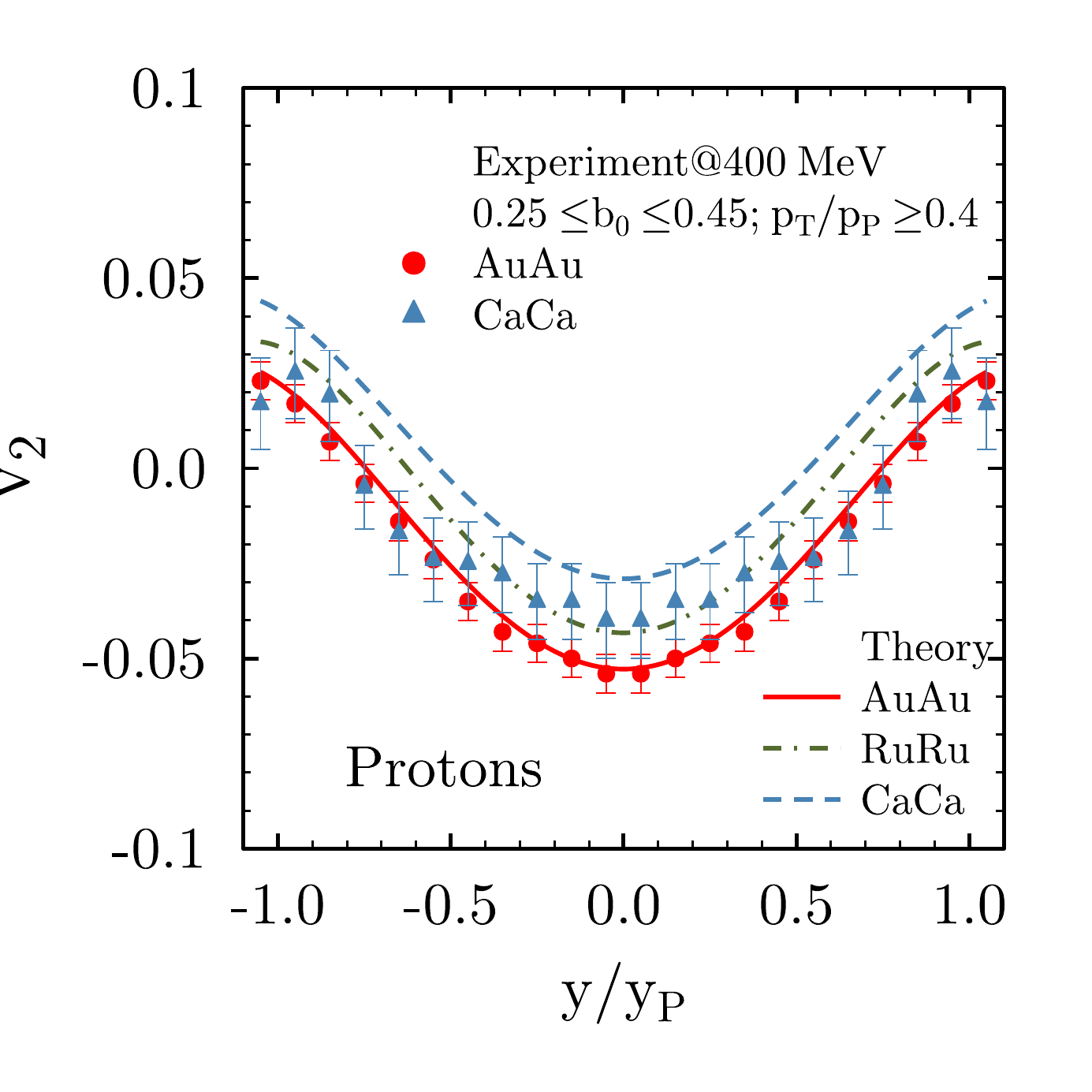}
 \caption{(Left Panel) Model dependence for elliptic flow of protons. Theoretical curves have the same meaning as those in the
 left panel of \figref{stopping}. (Right Panel) System size dependence of elliptic flow of protons. Experimental data are taken from Ref. ~\cite{FOPI:2011aa}.}
 \figlab{ellipticflow}
\end{figure*}

\subsection{Benchmarking the nucleonic sector}
\seclab{2c}

The time evolution of heavy-ion collisions at impact energies of a few hundred MeV/nucleon is governed by nucleonic degrees of
freedom. In order to realistically describe pion production at these energies it is crucial that nucleonic multiplicity spectra are accurately reproduced in order to have the correct invariant mass spectra of two-body collisions. To this end, before embarking on a study of pion production, a theoretical transport model would have to pass the test of comparing predictions for nucleonic observables to experimental data. In particular a proper description of stopping and flow observables is mandatory.

In a previous publication theoretical predictions for transverse and elliptic flows for {$^{197}$Au+{$^{197}$Au at an impact energy of 400 MeV/nucleon were compared to experimental FOPI data in the context of extracting constraints for the density dependence of the symmetry energy~\cite{Cozma:2017bre}. In this Section, the theory-experiment comparison is extended by investigating stopping and system size dependence of observables in the 150 to 1000 MeV/nucleon impact energy range.

In the left panel of \figref{stopping} theoretical predictions for the stopping observable $varxz$ of protons in central
{$^{197}$Au+{$^{197}$Au are presented and compared to experimental data ~\cite{Reisdorf:2010aa}. The impact
of relevant models ingredients is shown in order to assess model uncertainties. The full model predictions (full curve) describe low impact energy data very well. At the higher end of the incident energy interval a slight underprediction is however noticeable. For comparison, full model predictions employing a different Pauli blocking algorithm that estimates occupancy fractions making use of the Gaussian wave function associated to each nucleon rather than the standard TuQMD algorithm ~\cite{Cozma:2017bre} are presented (dashed curve). The difference is small at all incident energies. The importance of threshold effects and the multi-nucleon correlations they induce is underlined by comparing the predictions of the model with these effects switched off (dash-dotted curve) to the full model (full curve). Their impact is larger at lower incident energies, the difference between the two calculations amounting to about 10$\%$. The magnitude of the effect is surprising in view of the fact that shifts of the invariant mass of the colliding nucleons amounts to a few MeV~\cite{Cozma:2014yna}. At a basic level the effect is a consequence of stronger energy dependence of elastic collision and the nucleon optical potential at lower incident energies.

The impact of in-medium modifications of cross-sections is demonstrated by switching off these effects to inelastic channels and then additionally also to the elastic ones. As expected, in-medium modifications of inelastic cross-section affect stopping observables only above 500 MeV/nucleon impact energy. The Ansatz of relating these medium corrections to effective masses
induces an energy dependence of $varxz$ that deviates visibly from the experimental one even though the absolute magnitudes are still reproduced. At low impact energies, modifications of elastic nucleon-nucleon cross-sections are crucial to describe experimental data and a momentum dependence of these effects appears to be mandatory. Similar conclusions have been reached in other studies \cite{Barker:2016hqv,Wang:2013wca,Li:2011zzp}.

The same analysis has also been performed for deuteron and triton stopping in {$^{197}$Au+{$^{197}$Au collisions for which experimental measurement are available \cite{Reisdorf:2010aa}. The relevance of the above discussed model ingredients remains similar, however the overall description of the experimental data is poorer. Deuteron stopping is under-predicted by approximately 15$\%$, while for tritons the deviation increases to 35$\%$. This is not surprising in the context of triton multiplicities being under-estimated by a factor of about 2~\cite{Cozma:2017bre} by the model. Switching off in-medium effects on cross-sections reduces the discrepancy considerably but the induced energy dependence of the observable at the lower limit for the
impact energy is not realistic.

The right panel of \figref{stopping} presents predictions of the full model for proton stopping in central collisions
for three different systems: {$^{197}$Au+{$^{197}$Au, {$^{58}$Ni+{$^{58}$Ni and  {$^{40}$Ca+{$^{40}$Ca. Experimental results at
impact energies for which data are available~\cite{Reisdorf:2010aa} are also shown. A generally good agreement between theory and experiment is observed.

Turning to elliptic flow, in the left panel of \figref{ellipticflow} predictions for transverse momentum dependent elliptic flow of protons in {$^{197}$Au+{$^{197}$Au collision at an impact energy of 250 MeV/nucleon are presented. Similarly as for stopping, the impact of certain model ingredients is shown. Only in-medium modifications of elastic cross-sections lead to a significant departure from the full model predictions, while threshold effects and different approaches of computing the nucleon occupancy have a negligible impact. The full model is in almost in perfect agreement to the corresponding experimental data~\cite{FOPI:2011aa}. By comparing the left panels of \figref{stopping} and \figref{ellipticflow} it is evident that a simultaneous description of both stopping and elliptic flow is not possible by solely introducing in-medium modifications of elastic cross-sections. The inclusion
of threshold effects appears almost indispensable. As the incident energy is increased the impact of in-medium modifications of elastic cross-sections on elliptic flow decreases, a good description of the experimental data is still achieved \cite{Cozma:2017bre}. Investigation of elliptic flow of deuterons and tritons has lead to the same conclusions. 

The right panel of ~\figref{ellipticflow} presents predictions for rapidity dependent elliptic flow at an impact energy of 400 MeV/nucleon for three systems: $^{197}$Au+{$^{197}$Au, {$^{96}$Ru+{$^{96}$Ru and {$^{40}$Ca+{$^{40}$Ca. Experimental data are available only for the first and third systems~\cite{FOPI:2011aa}. An excellent description of $^{197}$Au+{$^{197}$Au data is observed, the strength of the predicted elliptic flow of protons for {$^{40}$Ca+{$^{40}$Ca collisions is slightly weaker than the experimental one. A similar picture is valid for the elliptic flow of deuterons for the same reactions.

A similar study has been performed for transverse flow. None of the model ingredients studied above have a significant impact for this observable and consequently the quality of the description of the experimental data is similar to that of Ref.~\cite{Cozma:2017bre} for all impact energies in the range of interest and for all light cluster species for which experimental data have been reported in Ref.~\cite{FOPI:2011aa}.

\begin{figure*}
\begin{center}
 \includegraphics[width=0.8\textwidth]{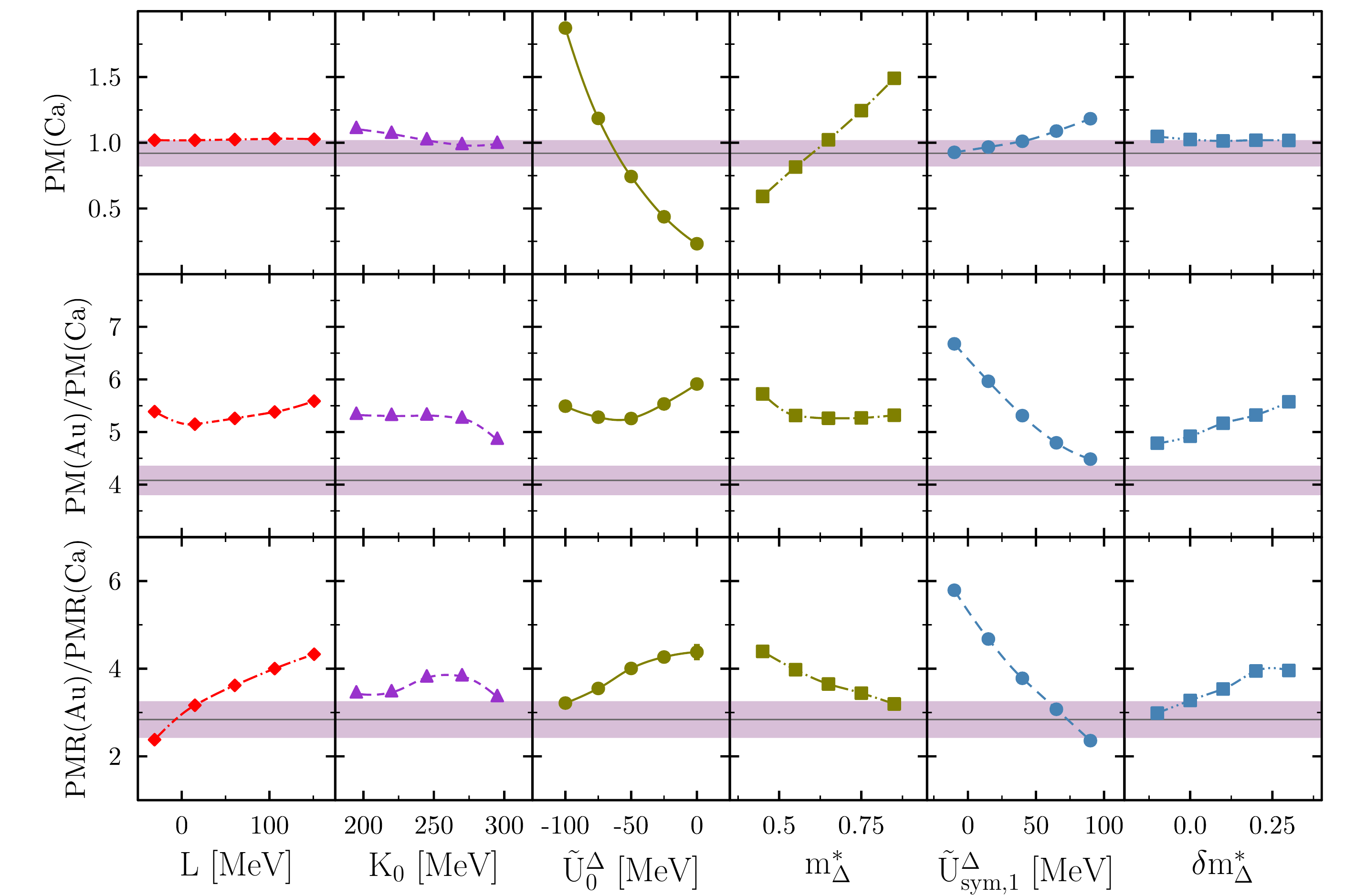}
 \end{center}
 \caption{Dependence of the total charged pion multiplicity in CaCa collisions (top panels), ratio of charged pion multiplicities of AuAu to CaCa (middle panels) and double charged pion multiplicity ratio of AuAu to CaCa (bottom panels) on $\Delta$ (1232) potential depths at saturation $\tilde U_{is}^{\Delta}$, $\tilde U_{iv}^{\Delta}$, effective mass parameters $m^*_\Delta$ and $\delta m^*_\Delta$, compressibility modulus of symmetric nuclear matter $K_0$ and slope of symmetry energy at saturation $L$ for central ($b_0<$0.15) collisions at 400 MeV/nucleon impact energy. The corresponding experimental values \cite{Reisdorf:2010aa} are depicted by horizontal bands.}
 \figlab{pionobs}
\end{figure*}

\section{Impact of the $\Delta$ (1232) potential on pionic observables}
\seclab{3}
The magnitudes of ISDP and IVDP are poorly known at best, as already emphasized in previous sections. It is thus mandatory to identify a sufficient number of observables to extract both the values for the parameters used to fix these potentials and those describing the density dependence of the EoS. Originally, the charged pion multiplicity ratio was proposed as an observable for extracting the value of the slope parameter $L$ of SE \cite{Li:2004cq}. %To determine the four parameters fixing the DPOT at least four additional observables are needed.
In a previous publication~\cite{Cozma:2016qej} the average transverse momentum of charged pions was used to constrain the strength of the IVDP relative to the nucleon symmetry potential. This observable has however been proven very sensitive to the pion optical potential. To avoid additional model dependence we will restrict the present study to multiplicity related observables only. An obvious choice is the total charged pion multiplicity, for which experimental data exist for several systems at various impact energies ~\cite{Reisdorf:2010aa}. Owing to the limited applicability of the approximations used in taking into account threshold effects~\cite{Cozma:2014yna} the upper limit of the impact energy will be restricted to 800 MeV/nucleon. Consequently the following available experimental data for given systems and impact energies in MeV/nucleon can be used: $^{40}$Ca$^{40}$Ca (400, 600, 800), $^{96}$Ru$^{96}$Ru (400), $^{96}$Zr$^{96}$Zr (400) and $^{197}$Au$^{197}$Au (400, 600, 800). These systems have the following average isospin asymmetry: 0.0 (CaCa), 0.08 (RuRu), 0.17 (ZrZr) and 0.20 (AuAu) allowing the study of both ISDP and IVDP. The rather broad range of impact energies will also facilitate the study of their momentum dependence. In the near future experimental data with significantly better accuracy for $^{108}$Sn$^{112}$Sn, $^{112}$Sn$^{124}$Sn and $^{132}$Sn$^{124}$Sn at an impact energy of 270 MeV/nucleon, slightly below threshold, will become available \cite{Shane:2014tsa} and potentially provide tighter constraints. 

\subsection{Relevant observables}
\seclab{3a}

For each system two independent observables can be constructed from charged pion multiplicities: total charged pion multiplicity (PM) and charged pion multiplicity ratio (PMR). For two systems at the same impact energy, one neutron rich and one neutron deficient, two observables, dependent on the two PMs and two PMRs can be defined: the ratio of total charged pion multiplicities and double charged pion multiplicity ratio. Each of them are useful in studying the impact of DPOT on pionic observables.

In the top panels of \figref{pionobs} the sensitivity of PM to DPOT parameters, compressibility modulus of symmetric nuclear matter and slope $L$ of the symmetry energy is presented. The standard choice for the six mentioned parameters is (see also \tabref{model_input_params}): $\tilde U_{0}^{\Delta}$=-67.0 MeV, $\tilde U_{sym,1}^{\Delta}$=45.0 MeV, $m^*_{\Delta}$=0.65, $\delta m^*_{\Delta}$=0.175, $K_0$=245 MeV and $L$=60 MeV. The values of the first four parameters lead to ISDP and IVDP that resemble closely the corresponding ones of nucleon. The calculations presented in the figure were performed by varying the indicated parameter within a reasonable interval, as represented by the abscissa of the corresponding plot, while the values of the other five parameters are kept unchanged to their standard ones. The PM displays considerable sensitivity to the isoscalar potential depth at saturation $\tilde U_{0}^{\Delta}$ and the value of the isoscalar effective mass $m^*_\Delta$. There is a comparably much smaller sensitivity, of the order of 10-20$\%$, to the strength of IVDP at saturation $\tilde U_{sym,1}^{\Delta}$ and to the value of the compressibility modulus. The sensitivities to the remaining two parameters, isovector mass difference $\delta m^*_{\Delta}$ and slope parameter $L$ are negligibly small. The sensitivity to $\tilde U_{sym,1}^{\Delta}$ is small, but not negligible, for the $^{40}$Ca$^{40}$Ca system. In fact the smallest sensitivity to this parameter was found for $^{96}$Ru$^{96}$Ru. The slope of the dependence of PM on $\tilde U_{sym,1}^{\Delta}$ for $^{197}$Au$^{197}$Au has an opposite sign to that derived from the top panel of \figref{pionobs} for $^{40}$Ca$^{40}$Ca. This suggests that the net effect is the result of two opposite trends related to the average isospin asymmetry and fluctuations. It is concluded that PM is suitable to fix the parameters of ISDP. Secondary order corrections due to IVDP are not negligible and will have to be accounted for.

In the middle panels of \figref{pionobs} a similar analysis is presented for the ratio of total charged pion multiplicities of $^{197}$Au$^{197}$Au to $^{40}$Ca$^{40}$Ca at impact energy of 400 MeV/A. The important feature of this observable is that the huge dependence on ISDP evidenced for total pion multiplicities of individual systems almost cancel out, with a remaining residual sensitivity of about 10$\%$. The dominant variations are related to the isovector potential depth at saturation $\tilde U_{sym,1}^\Delta$ and isovector effective mass difference $\delta m^*_{\Delta}$. Sensitivity to the equation of state parameters $K_0$ and $L$ is also in this case limited to about 10$\%$. Constraining the isovector potential parameters using this observable appears feasible but model dependence is not negligible due to relatively important sensitivity to other parameters. Using the standard values for model parameters the calculation overestimates the experiment by 30$\%$. Varying IVDP parameters within a conservative interval does not allow a mitigation of the discrepancy. An investigation of this issue suggests that a possible resolution may involve modifications of in-medium $\Delta$ production cross-sections, a stiff density dependence of IVDP above saturation (see below) or a larger positive value for the neutron-proton effective mass difference (the standard choice being 0.33$\beta$).

The third sensitivity study was performed for the double charged pion multiplicity ratio of $^{197}$Au$^{197}$Au to $^{40}$Ca$^{40}$Ca at 400 MeV/nucleon impact energy. The results are presented in bottom panels \figref{pionobs} for the same model parameters as above. For this observable the sensitivity to each of the chosen model parameters is sizable. This is the only observable of the three that shows sizable dependence to asy-EoS. However, the extraction of the value of $L$ is impeded by the unknown values of DPOT parameters. Setting this quantity equal to that of the nucleon has been in the past a choice of convenience that generally does not lead to a good description of all available experimental data. The alternative approach of modifying in-medium $\Delta$ production cross-section is restricted by existing microscopical models \cite{Larionov:2003av} that suggest that such effects are largely governed by scaling laws involving in-medium effective masses. Such effective modifications of inelastic cross-sections have been included in the present model.

\begin{figure}
 \includegraphics[width=0.495\textwidth]{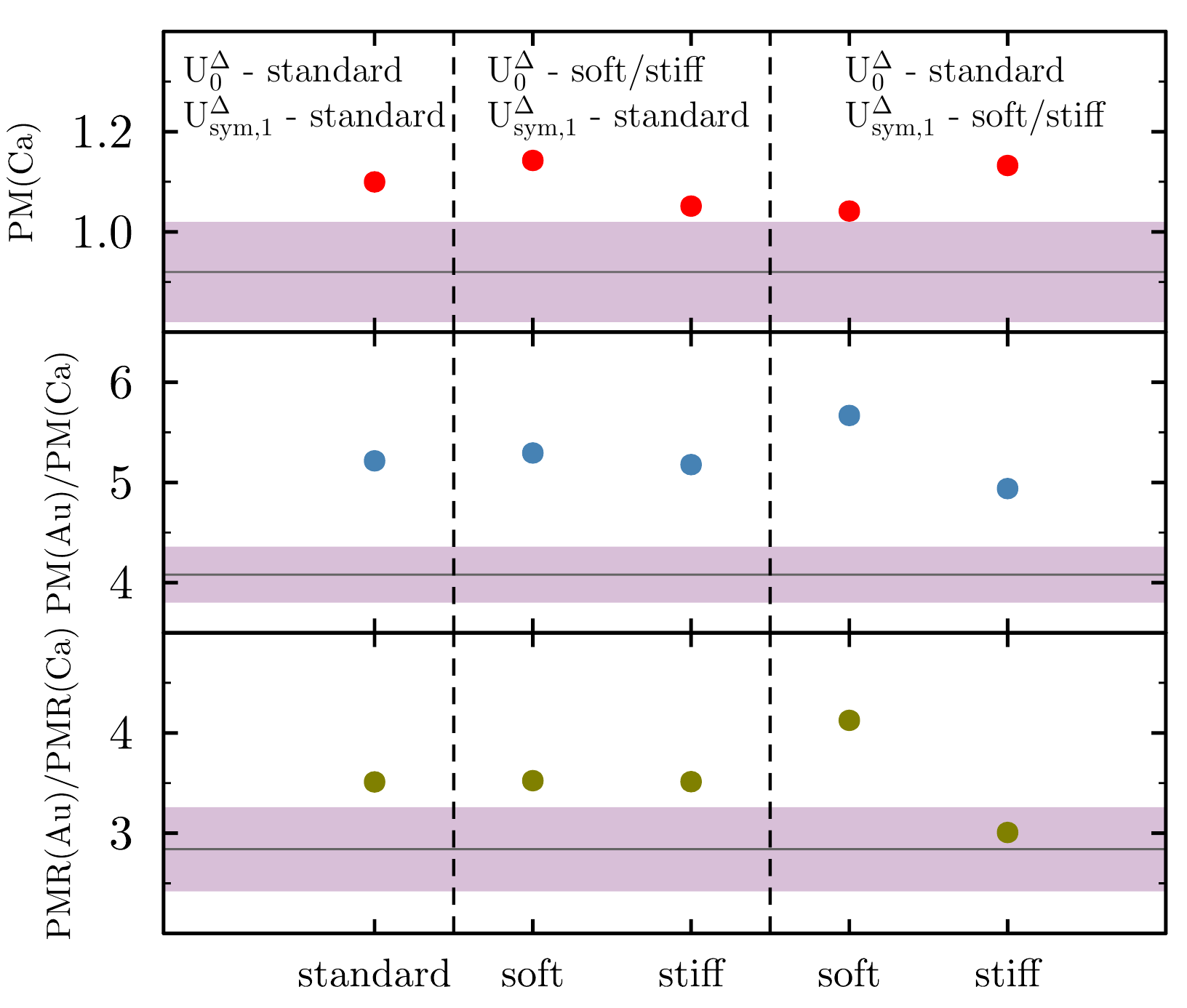}
 \caption{Dependence of the total charged pion multiplicity in CaCa collisions (top panel), ratio of charged pion multiplicities of AuAu to CaCa (middle panel) and double charged pion multiplicity ratio of AuAu to CaCa (bottom panel) on the density dependence of $U_0^\Delta$ and $U_{sym,1}^\Delta$ potentials above saturation for central ($b_0<$0.15) collisions at 400 MeV/nucleon impact energy. Results using the standard choice for the supranormal density dependence of these potentials are presented in the leftmost column for comparison. The corresponding experimental values \cite{Reisdorf:2010aa} are depicted by horizontal bands.}
 \figlab{pionobsdd}
\end{figure}

In \secref{2b} soft and stiff density dependent ISDPs and IVDPs have been constructed and displayed in \figref{iscpot} and \figref{isvpot} respectively. In \figref{pionobsdd} their impact on the three observables discussed above is presented and compared with predictions for the ``standard'' case potential whose parameters values are listed in \tabref{model_input_params}. Modifying the density dependence of ISDP above saturation has an impact on total charged pion multiplicities (PM) of at most 10$\%$, a softer density dependence of the potential leading to an increase of total pion multiplicities. The ratio of total charged pion multiplicities and the double charged pion multiplicities ratio of AuAu to CaCa at 400 MeV/nucleon are only marginally affected by a soft/stiff density dependence of ISDP above saturation. Modification of the density dependence of the IVDP has a visible impact on all three observables. PM is affected at 5$\%$ level, the impact on PM(Au)/PM(Ca) is close to $10\%$ and the effect on the double ratio PMR(Au)/PMR(Ca) is the largest at 20$\%$. Comparison with \figref{pionobs} reveals that the sensitivity to the density dependence of IVDP above saturation is several times smaller than to the magnitude of the symmetry potential at saturation $\tilde U_{sym,1}^\Delta$. The same observation is true also for ISDP by an even larger margin. This provides an a posteriori justification for the choice of parameters used in this study to fix the density dependence of DPOT: its strength at saturation and at twice saturation density.

Sensitivity of each of the three observables has also been studied with respect to the following ingredients of the transport
model: pion potential, neutron-proton effective mass difference and in-medium modification of both elastic and inelastic cross-sections. The impact is found to lie in the interval 5-10$\%$ in each case, with the exception of medium modifications of cross-sections that impact total pion multiplicities at the level of 20$\%$ for light systems such as CaCa. It is concluded that the three observables can be used to extract information on the strength of the ISDP and IVDP at saturation $\tilde U_{0}^\Delta$ and $\tilde U_{sym,1}^\Delta$ and the isoscalar effective mass of $\Delta$(1232) in nuclear matter $m^*_\Delta$. Extracting constraints for $\delta m_\Delta^*$ is feasible but dependence of results on other model parameters (such as the slope parameter of asy-EoS and the density dependence of IVDP above saturation) cannot be neglected.

Constraints for the density dependence of DPOT above saturation are necessary for a complete knowledge of this quantity. The results in this Section do however prove that this is not feasible at present given the small impact on the studied observables, which is comparable or even smaller than the uncertainties of other not precisely known model ingredients such as pion potentials or in-medium cross-sections. Consequently, in the following we will only present the impact of the high density dependence of DPOT on constraints for $\tilde{U}_{0}^\Delta$, $\tilde{U}_{sym,1}^\Delta$, $m_\Delta^*$ and $\delta m_\Delta^*$ rather than attempting to extract values for $U_{0}^\Delta$ and $U_{sym,1}^\Delta$ at 2$\rho_0$.

A similar study has been performed for momentum related observables. Also in this case the impact of the DPOT
is sizable but of comparable relative magnitude to that of the pion optical potential. Using such observables would induce important model dependence of results, as already evidenced in Ref. ~\cite{Cozma:2016qej}, and will not be pursued here.
%\section{Relevant observables}
%\label{sec:3a}

\subsection{Constraining $\Delta$ (1232) potential parameters}
\seclab{3b}

Using the insights of the previous Section we proceed to extract constraints for the values of DPOT parameters. Results for the isoscalar component are presented in \figref{fitiscpot} as correlations between the isoscalar potential
depth at saturation $\tilde{U}_{0}^\Delta$ and isoscalar effective mass $m_{\Delta}^*$. To fix this quantity the available experimental data comprise those of the following (nearly) isospin symmetric systems: CaCa at 400, 600 and 800 MeV/nucleon and RuRu at 400 MeV/nucleon central collisions \cite{Reisdorf:2010aa}. In the left panel constraints for ISDP parameters extracted from collisions of CaCa at 400 MeV/nucleon are presented in the form of 1-$\sigma$ confidence level contour plots. Besides a calculation employing the full model, certain model ingredients have been modified or switched off to test model dependence. Three additional simulations correspond to the full model making use of a Pauli blocking algorithm based on computation of the occupancy fraction using the Gaussian wave function associated to each nucleon~\cite{Cozma:2017bre}, full model without the pion potential and full model with vacuum inelastic cross-sections. The first two lead to results compatible with the full model, while for the third the deviation is more important as a consequence of total pion multiplicities being impacted at 20$\%$ level by in-medium modifications of inelastic cross-sections. For heavier systems (such as AuAu) or light systems at higher impact energies the effect
of medium modification of inelastic cross-sections on multiplicities is smaller, in the 10-15 $\%$ range. Additionally, constraints extracted using the soft and stiff density dependent $U_0^\Delta$ potentials above saturation, introduced in \secref{2b}, are also presented. The impact of modifying the strength of $U_0^\Delta$ at 2$\rho_0$ is small, similar in magnitude to the effect due to the pion potential. Calculations using a soft/stiff density dependence of $U_{sym,1}^\Delta$ are not shown, however results in \figref{pionobsdd} allow the inference that their impact is of similar small magnitude as for $U_0^\Delta$. Consequently the PM observable can only be used to study the $U_0^\Delta$ potential close or below saturation by determining values for $\tilde{U}_0^\Delta$ and $m_\Delta^*$. 

\begin{figure*}
 \includegraphics[width=0.495\textwidth]{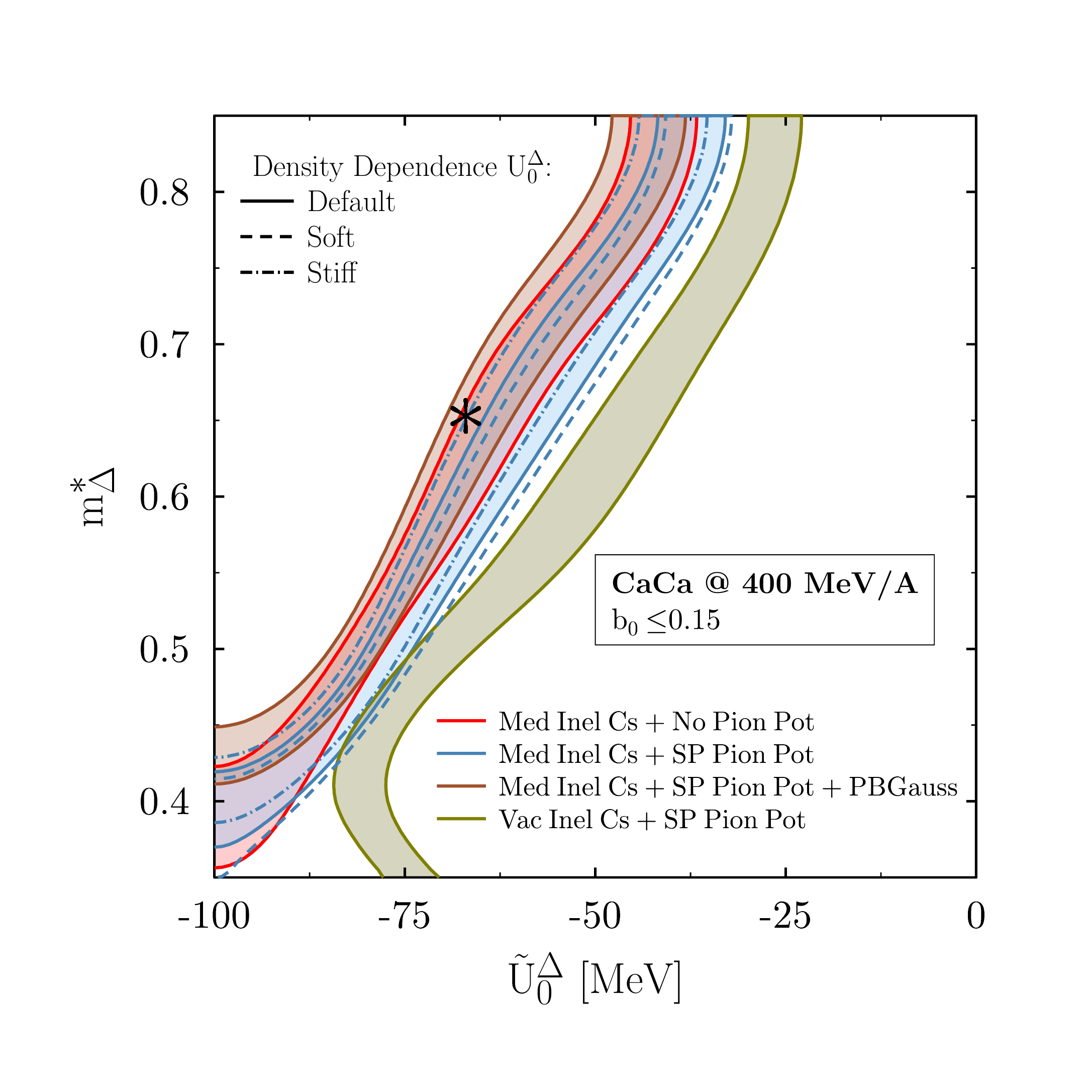}	
 \includegraphics[width=0.495\textwidth]{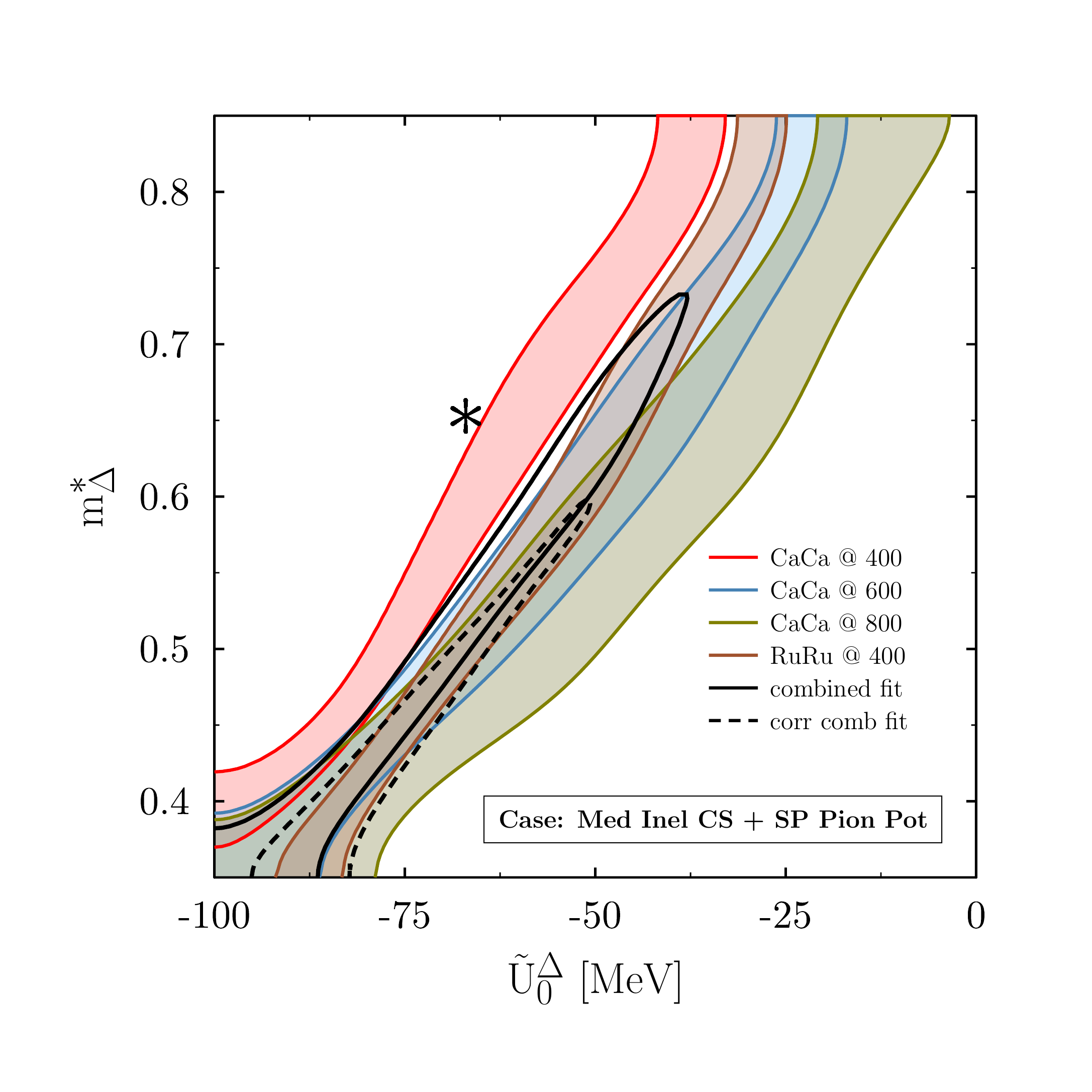}
 \caption{(Left Panel) Model dependence of constraining ISDP depth
 at saturation density and zero momentum $\tilde{U}_{0}^\Delta$ and isoscalar effective mass $m_{\Delta}^*$ for central CaCa
 collisions at 400 MeV/nucleon. Results for six different cases are shown: full model, different Pauli blocking algorithm, no pion potential, no in-medium effects on inelastic channels cross-sections and soft/stiff density dependence of $U_0^\Delta$ above saturation. The star represents the choice of potential parameters that would render the ISDP equal to nucleon's. (Right Panel) Constraints for the same parameters by making use of the FOPI experimental data \cite{Reisdorf:2010aa} for central collisions of RuRu at 400 MeV/nucleon and CaCa at 400, 600 and 800 MeV/nucleon. The combined result for the four reactions is represented by the contour plot labeled ``combined fit''. For the contour plot labeled ``corr comb fit'' corrections due to sensitivity to IVDP have been applied, as described in the text. All contour plots correspond to 1-$\sigma$ confidence level of fitting total charged pion multiplicity.}
 \figlab{fitiscpot}
\end{figure*}

In the attempt to pin down the momentum dependence of ISDP, simulations of collisions at different impact energies have been performed and compared to experimental data. Constraints for $\tilde{U}_{0}^\Delta$ and $m_{\Delta}^*$ are presented in the right panel of \figref{fitiscpot}. Simulations for RuRu at 400 MeV/nucleon have also been performed and since total multiplicity for this system has displayed the smallest sensitivity to the isovector component of the potential, as previously mentioned, it has been added to the comparison. It is evident that the 1-$\sigma$ CL contour plots for CaCa at different impact energies have slightly different slopes in the ($\tilde{U}_0^\Delta$,$\,m_{\Delta}^*$) plane, converging at smaller values for $m_{\Delta}^*$. A combined fit of the
four reactions is sub-optimal with a minimum value of $\chi^2$/point=2.55. As has been evidenced in the upper panels of \figref{pionobs} there is still non-negligible residual dependence on the IVDP strength. It was found that it affects
total pion multiplicities of CaCa by 15, 10 and 5 $\%$ at 400, 600 and 800 MeV/nucleon impact energies respectively. Once this is taken into account a corrected combined fit with a minimum value of $\chi^2$/point=0.80 is obtained. The combined fit of the four reactions has somewhat restricted the possible values for $\tilde{U}_{0}^\Delta$ and $m_{\Delta}^*$, definite values could however not be extracted in part because the experimental data carry rather large uncertainties but also because at higher impact energies the sensitivity decreases. Near-future availability of experimental data slightly below pion production threshold for the nearly isospin symmetric $^{108}$Sn$^{112}$Sn system by the SPIRIT collaboration may improve the present situation significantly. In ~\figref{fitiscpot} the values of parameters leading to an ISDP equal to that of the nucleon are depicted by a star symbol. It departs from the favored parameter values of the combined fit for the four systems at more than 5-$\sigma$ CL. However, within this approach it is not clear whether this is a model independent conclusion as the favored values for the potential parameters may be the result of the fit compensating for some drastic approximations.

%O'Connell:1990zg,Bodek:2020wbk,deJong:1992wm,Baldo:1994fk
In Ref. \cite{Bodek:2020wbk} the strength of the DPOT was extracted from the study of experimental data of quasi-elastic scattering of electrons on bound nucleons in nuclei of different masses: $^6$Li, $^{12}$C, $^{27}$Al, $^{40}$Ca/Ar and $^{56}$Fe. It was found that the DPOT is more attractive than the empirical nucleon optical potential and the attraction is stronger for heavier nuclei reflecting higher probed densities. The attraction is stronger by about 20 MeV at momenta close to $p$=0.5 GeV/c for the heaviest nuclei for which the analyses was performed. Additionally, an arguably stronger energy dependence was evidenced for momenta significantly above the Fermi sea, which may suggest a lower isoscalar effective mass of $\Delta$(1232). A previous similar study \cite{O'Connell:1990zg} has reached similar conclusions. These qualitative results are in full agreement with findings of the present study for the ISDP, as shown in the right panel of \figref{fitiscpot} due to a similar similar approach. 

Comparison with microscopic calculations reveals important differences. Many-body calculations of pion-nucleus scattering or absorption performed in the framework of the Delta-hole model arrived at the conclusion of an ISDP less attractive than that of the nucleon at saturation density: $\tilde{U}_{0}^\Delta \approx $-30 MeV ~\cite{Hirata:1977hg,Horikawa:1980cv,Oset:1987re,GarciaRecio:1989xa}. Contributions such as non-resonant background pion production, the spin-orbit component of the $\Delta$ potential and short-range corrections to interaction vertices are crucial for the quoted result. From the upper panels of \figref{pionobs} it is evident that by inclusion of non-resonant background contributions to pion production in the scattering term of the transport model a less attractive ISDP will be favored. Ab-initio calculations, that have used well established microscopical nucleon-nucleon potentials (such as Argonne $v_{28}$) as input, performed within the framework of the Bethe-Brueckner-Goldstone method~\cite{Baldo:1994fk} or one-boson exchange nucleon-nucleon potentials in the relativistic Dirac-Brueckner model allowing a good reproduction of the elastic pion-nucleon $P_{33}$ phase-shift \cite{deJong:1992wm}, have arrived at a mildly repulsive ISDP. This is in part due to dominant repulsive contributions of total isospin I=2, a channel which cannot be sufficiently constrained by elastic nucleon-nucleon scattering data.

To proceed to extraction of constraints for IVDP parameters $\tilde{U}_{sym,1}^\Delta$ and $\delta m_{\Delta}^*$ specific choices need to be made for ISDP parameters. The choice $\tilde{U}_0^\Delta$=-78 MeV and $m_{\Delta}^*$=0.45 allows, as evidenced in the right panel of \figref{fitiscpot}, a good description of pion multiplicity for isospin symmetric (or nearly so) systems. In \figref{fitisvpot} the favored values for $\tilde{U}_{sym,1}^\Delta$ and $\delta m_{\Delta}^*$, resulting from comparing theoretical and experimental values of PMR for central AuAu collisions at 400 MeV/nucleon incident energy, are shown as 1-$\sigma$ CL contour plots. Results for three different values of the slope parameter $L$ of SE are shown, evidencing an important dependence on this parameter. Additionally, the soft and stiff density dependent $U_{sym,1}^\Delta$ introduced in \secref{2b} lead to differences in the extracted constraints of similar magnitude as those induced by $L$. It has been verified that the extracted constraints are sensitive also to the value of the neutron-proton effective mass difference. Results obtained by fitting the experimental value of PMR for ZrZr central collisions at 400 MeV/nucleon are also shown for $L$=60 MeV and the standard density dependence above saturation. They are compatible with the corresponding ones for AuAu. Adding the total multiplicity of charged pions for these isospin asymmetric systems in the fit, slightly restricts the allowed ranges, by disfavoring regions of higher values for $\tilde{U}_{sym,1}^\Delta$. The star symbol represents the choice for the potential parameters that would lead to an identical isovector potential for nucleons and $\Delta$(1232) isobars. Constraints extracted for a widely used value of the slope parameter, $L$=60 MeV, depart from this commonly made choice, but by a smaller margin compared to the isoscalar case. 

Fitting available experimental data of PMR for AuAu at higher impact energy does not bring additional information mainly due to the larger experimental error for this observable. The second observable of interest for constraining the isovector
$\Delta$(1232) potential, the ratio of total charged-pion multiplicities, has proven ineffective, nearly half of the probed parameter space in \figref{fitisvpot} leading to theoretical predictions in accord to experiment.

It becomes clear that a unique extraction of DPOT using multiplicity observables alone is not possible at present. In principle, the analysis can be extended to include existing information related to momenta of pions. Published results for the ratio of average $p_T$ of charged pions exists in the literature \cite{Reisdorf:2010aa} and have been used for this
purpose in the past~\cite{Cozma:2016qej}. The additional induced model dependence from the isoscalar channel, as shown in the left panel of \figref{fitiscpot}, is not negligible and the extraction of the stiffness of SE will carry an even larger model dependence. Determining the slope of the SE is in principle possible by performing a five parameter fit of multiplicity and momentum related observables. This avenue has been explored. The resulting value for $L$ does however carry large uncertainties.

\begin{figure}
  \includegraphics[width=0.45\textwidth]{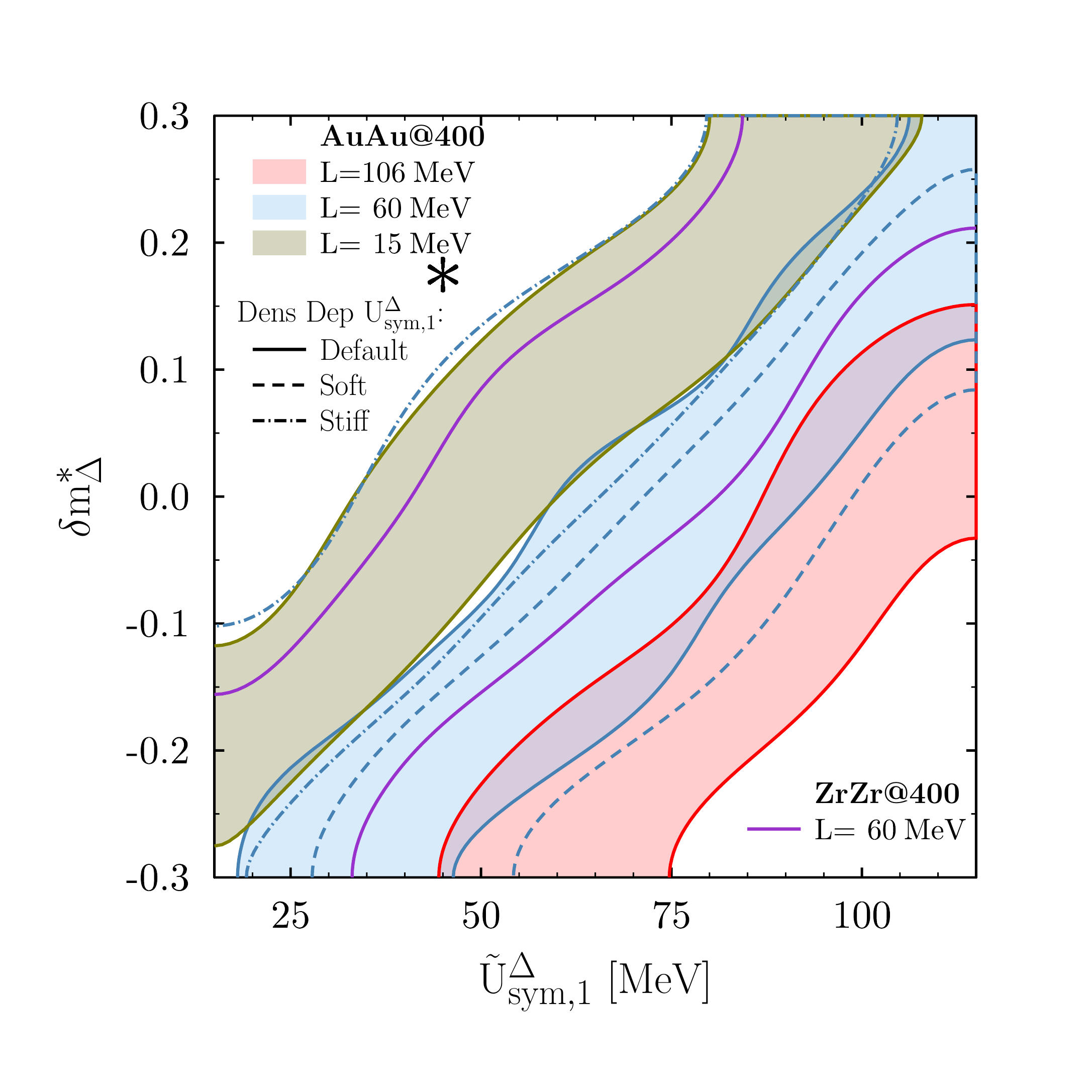}	
  \caption{Constraints for the IVDP parameters, potential depth at saturation density
  and zero momentum $\tilde{U}_{sym,1}^\Delta$ and isovector mass difference $\delta m_{\Delta}^*$, extracted from FOPI experimental data \cite{Reisdorf:2010aa} for PMR in central ZrZr and AuAu collisions at 400 MeV/nucleon. The star represents the choice of potential parameters that would render IVDP equal to nucleon's. All contour plots correspond to 1-$\sigma$ confidence level.} 
  \figlab{fitisvpot}
\end{figure}

\begin{figure}
  \includegraphics[width=0.45\textwidth]{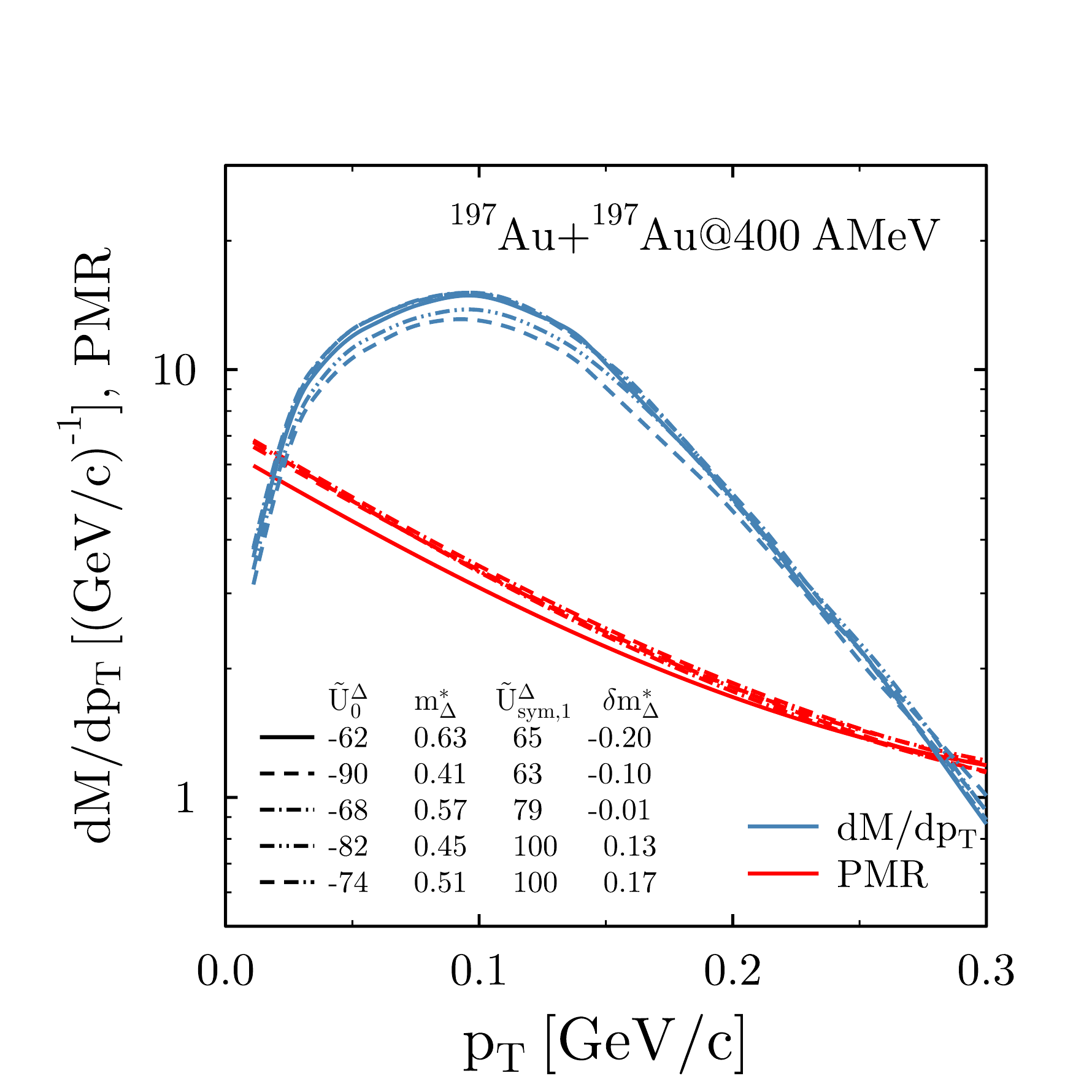}	
  \caption{Numerical proof that any choice for DPOT parameters, corresponding to both the isoscalar and isovector potentials, that lie on the 2-dimensional subspace of the 4-dimensional parameter space that results from fitting experimental total charged pion multiplicities and pion multiplicity ratios lead to simulated spectra are the same up to uncertainties induced by experimental data uncertainties. Parameter values for five such choices together with the theoretical
  multiplicity and PMR $p_T$ spectra are shown for mid-central (0.25$\leq b_0 \leq $0.45) AuAu collisions at 400 MeV/nucleon.The standard values for the parameters determining the density dependence the EoS of nuclear matter are used ($K_0$=245 MeV and $L$=60 MeV).} 
  \figlab{ptratspectra}
\end{figure}

Close to the pion production threshold it is possible to partially avoid the issue of the not uniquely extracted DPOT.
By fitting experimental multiplicities the 4-dimensional parameter space fixing DPOT at saturation is projected onto a 2-dimensional subspace. For any choice of the remaining two unconstrained parameters pion spectra are almost identical. This is a consequence of the fact
that close to threshold $\Delta$ degrees of freedom have no impact on the time evolution of the reaction in view of their scarcity. Consequently, nucleon spectra, which determine the distribution of invariant masses at which inelastic collisions take place, remain for all practical purposes unaffected by the depth or momentum dependence of DPOT. At these energies the DPOT plays the role of normalization constants (zeroth order moments) for the spectra, allowing for a reduction of model dependence of higher order moments. Fitting pion multiplicities will thus preserve any asy-EoS dependence of these quantities. The situation at higher impact energies, close to 1 GeV/nucleon and above, where the fraction of nucleons excited to resonances in the high density fireball is non-negligible \cite{Ehehalt:1993cx}, is different.

Results of numerical calculations, that prove invariance of spectra to arbitrary choices of model parameters in the 2-dimensional subspace left unconstrained after experimental multiplicities have been fitted, are presented in \figref{ptratspectra}. A four dimensional fit for PM and PMR for mid-central AuAu collisions at 400 MeV/nucleon has been performed. Five combinations of parameter values for $\tilde{U}_0^\Delta$, $m_{\Delta}^*$, $\tilde{U}_{sym,1}^\Delta$ and $\delta m_{\Delta}^*$ for which the fit is perfect ($\chi ^2$/point=0.0) have been chosen such that the sets of values are diverse. These choices are listed in the legend of \figref{ptratspectra}. The resulting total multiplicity and pion multiplicity ratio spectra as a function of the transverse momentum $p_T$ are shown. They are definitely close to each other though not identical. Differences are due to experimental accuracy of these observables that were used to compute the value of $\chi^2$/point and to the interpolation in a 4-dimensional parameter space using a very limited number of points (3 for each dimension) spanning a rather large parameter space. Nevertheless, the spectra are for the majority of cases within a few percent of each other. Means of improving this numerical proof are obvious and with predictable results. The standard density dependence of $U_0^\Delta$ and $U_{sym,1}^\Delta$ above saturation has been used. Extending the fit to a 6-dimensional one, thus including two additional parameters that can be used to change $U_0^\Delta(2\rho_0,p=0)$ and $U_{sym,1}^\Delta(2\rho_0,p=0)$ will lead quantitatively to the same multiplicity and single ratio spectra.

In practice the following approach will be used. Two parameters of the DPOT, $m_{\Delta}^*$ and $\delta m_{\Delta}^*$, will be set to well chosen values. The remaining two, $\tilde{U}_{0}^\Delta$ and $\tilde{U}_{sym,1}^\Delta$ will be
uniquely determined from a fit to experimental data for PM and PMR. It should be stressed that such a procedure destroys the
predictive power of the model. The determined set of parameters can only be used for the particular combination of systems and impact energies used in the fit. No firm conclusions can be drawn from a possible description (or failure to do so) of a different system. In the next Section the choice $m_{\Delta}^*$=0.45 and $\delta m_{\Delta}^*$=0.0 will be used.

\begin{figure*}
\begin{center}
  \includegraphics[width=0.45\textwidth]{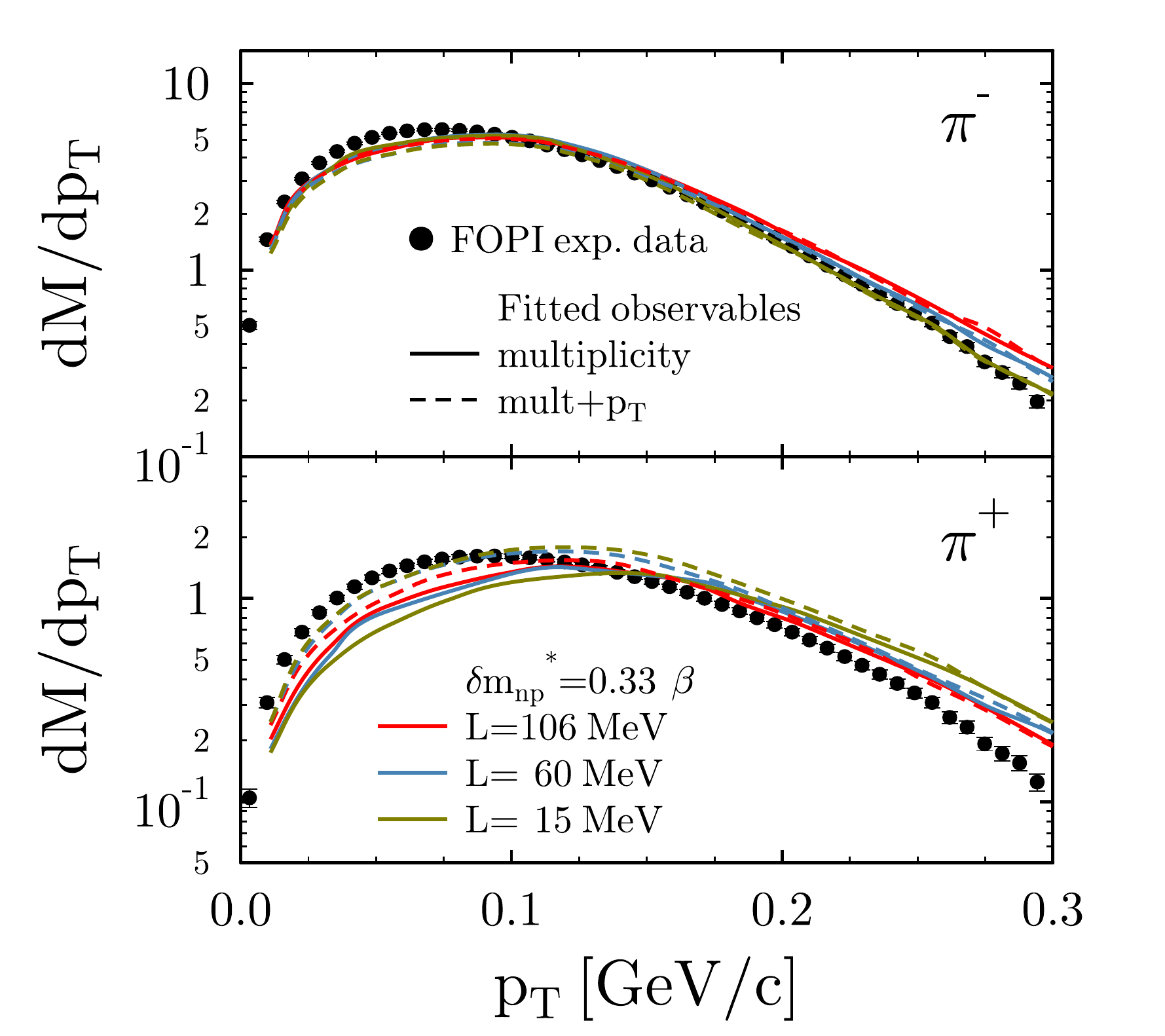}	
  \includegraphics[width=0.45\textwidth]{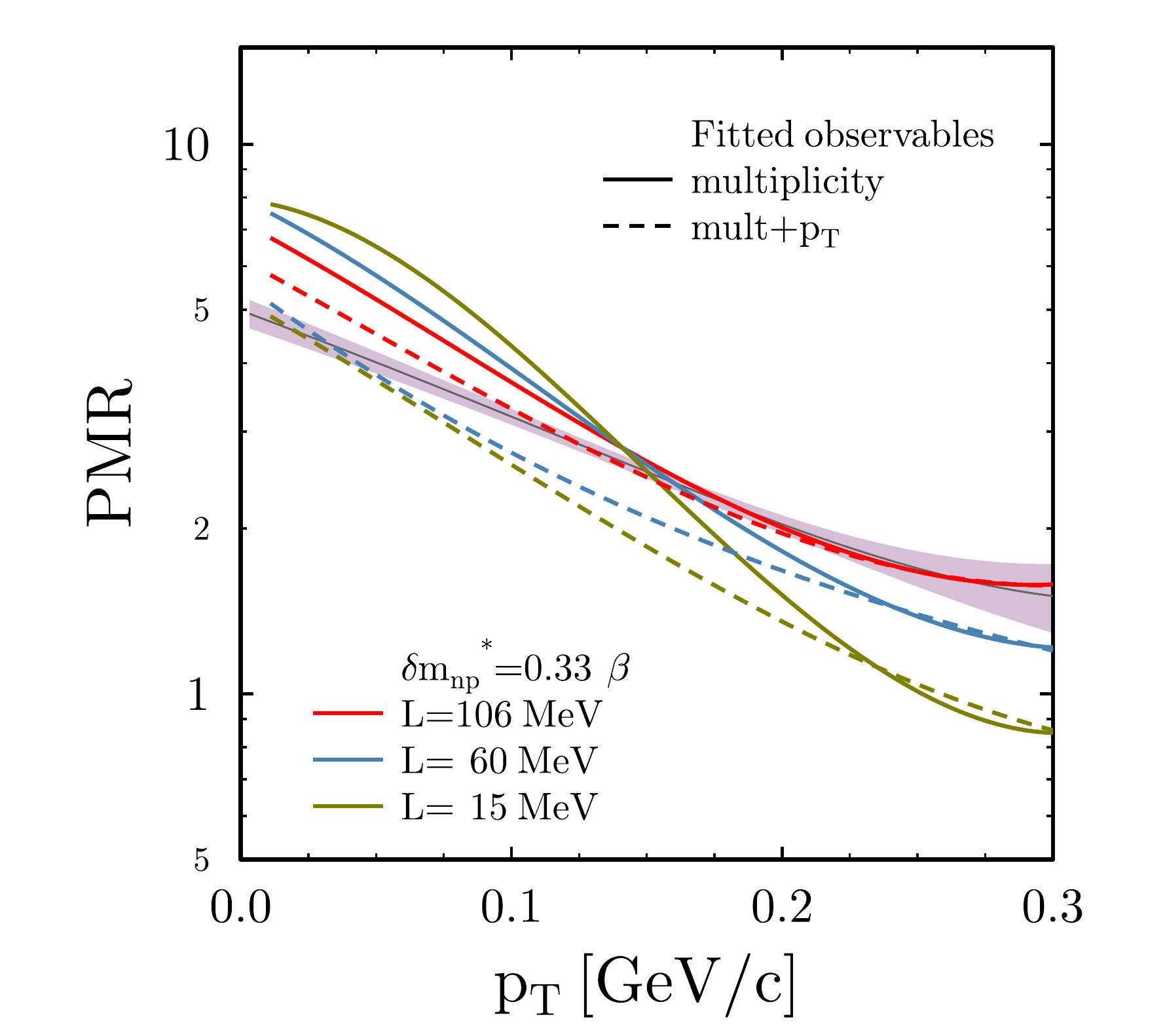}	
  \end{center}
  \caption{(Left Panel) Pion multiplicity and (Right Panel) pion multiplicity ratio $p_T$ dependent spectra obtained after fitting DPOT parameters to multiplicity observables for three values of the SE slope parameter $L$ and $\delta m_{np}^*$=0.33$\,\beta$. For each value of $L$ the impact on spectra of adding to the fitted observables also transverse momentum related ones (average combined transverse momentum and average $p_T$ ratio) is also shown. The simulations correspond to mid-central AuAu collisions at 400 MeV/nucleon impact energy. Unpublished FOPI experimental data \cite{Reisdorf:2015aa} are represented by full circle symbols (left panel) and band (right panel), the shown uncertainties being statistical.} 
  \figlab{multvsavptfit}
\end{figure*}

\section{Feasibility of constraining the symmetry energy}
\seclab{4}
\subsection{Sensitivity to asy-EoS and model dependence}
\seclab{4a}

In this Section the sensitivity of transverse momentum PMR spectra to the density dependence of SE and other relevant model parameters is studied. For this purpose simulations for mid-central ($0.25<b_0<0.45$) AuAu collisions at an incident energy of 400 MeV/nucleon have been performed. The sole motivation for the choice of this system has been the availability of experimental data.
Nevertheless, this data set is preliminary and does not account for systematical uncertainties, only statistical uncertainties being depicted for experimental data in figures of this Section. They have been employed in previous similar studies \cite{Song:2015hua,Cozma:2016qej} and can be useful in estimating the feasibility of studying the symmetry energy using this observable and the accuracy of the transport model. 

In \figref{multvsavptfit} a comparison of model prediction with experimental $p_T$ dependent individual pion multiplicity (left panel) and PMR (right panel) spectra is presented. One set of calculations (full curves) correspond to DPOT parameters extracted from a fit to PM and PMR observables, as described in \secref{3b}. Simulations for which DPOT parameters have been determined from a fit of both multiplicity and average transverse momentum observables are also displayed (dashed curves). The fitted momentum observables are: average transverse momentum of charged pions $\langle p_T^c \rangle=\frac{M_{\pi^-}\,\langle p_T^{\pi^-}\rangle+M_{\pi^+}\,\langle p_T^{\pi^+}\rangle}{M_{\pi^-}+M_{\pi^+}}$ and average transverse momentum ratio $\frac{\langle p_T^{\pi^+}\rangle}{\langle p_T^{\pi^-}\rangle}$. Each observable contributes to the total $\chi^2$ with the same weight. For each set, calculations for three values of the slope parameter of SE are provided: $L$=15, 60 and 106 MeV. The value for the neutron-proton effective mass difference has been set to its default value $\delta m_{np}^*$=0.33$\beta$.

The left panel of \figref{multvsavptfit} presents calculations for $\pi^-$ (top plot) and $\pi^+$ (bottom plot) multiplicity spectra.
Differences between the two sets of calculations are largest for $\pi^+$ spectra at low and intermediate $p_T$. Theoretical predictions for $\pi^+$ spectra are seen to deviate by important margins from experimental data at large values of $p_T$. Uncertainties in other model parameters, such as the neutron-proton effective mass difference $\delta m^*_{np}$, may explain this discrepancy (see below). Finer tuning of the symmetric part of EoS, in particular the compressibility modulus $K_0$ and nucleon isoscalar effective mass, preserving a consistent description of nucleonic observables, may also improve the description at high $p_T$ spectra of both $\pi^-$ and $\pi^+$ mesons.

 In the right panel of ~\figref{multvsavptfit} the PMR $p_T$ dependent spectrum is presented. Theoretical predictions display sensitivity to $L$ for all values of $p_T$, in relative terms they are largest at higher transverse momenta where predictions for the stiffest and softest choices of asy-EoS differ by a factor of almost 2. At higher $p_T$ values the two sets of predictions are nearly identical suggesting that this range of transverse momenta is free of model dependence originating in left-over uncertainties of the DPOT. The two sets of calculations become similar to each other for values of $p_T$ for which the strength in the multiplicity spectrum is below 10$\%$ of its peak value. 

The inclusion of momentum observables in the fit does not allow for a perfect fit, $Min(\chi^2)=$0, to be obtained anymore, but the the minimum of the merit function depends on other model parameters such as $L$ or the strength of the optical pion potential. The quality of the fit when momentum observables are included can in principle be improved by also varying $m_\Delta^*$ and $\delta m_\Delta^*$, rather than using the values mentioned at the end of \secref{3b}. In practice the discrepancy between model and experiment at low/moderate $p_T$ is reduced only modestly at the expense of performing a 4-dimensional fit (explicit calculations have been performed for the $L$=60 MeV case). The reason lies in the fact that to describe the spectra, moments of $p_T$ multiplicity distributions of order larger or equal to 1 need to be described by the model. The performed 4-dimensional fit only ensures that the 0$^{th}$ order moments are reproduced. In principle this approach can be used to constrain the asy-EoS parameters, but given the strong dependence of momentum observables on pion optical potentials, constraints extracted in this manner are rather imprecise~\cite{Cozma:2016qej}, as already argued in the previous Section. The high $p_T$ region appears thus better suited for studies of the SE. This will become clearer after other sources of model dependence of predictions in this region will be addressed below.

The sensitivity of PMR spectra to other two parameters of the EoS, \npEMD and value of SE at $\rho$=0.10 fm$^{-3}$, is presented in \figref{efmisvdependence}. The former quantity is varied within a range that includes most constraints for its value available in the literature: -0.33$\,\beta <\delta m_{np}^* < 0.66\,\beta$ \cite{Li:2018lpy}. Theoretical calculations reveal that the sensitivity to this parameter is almost as large as to the slope parameter of SE. This is a hardly surprising result since a different momentum dependence of the interaction results in different magnitudes of threshold shifts which in turn have been previously shown to have a large impact on PMR \cite{Song:2015hua,Cozma:2014yna}. To our best knowledge the impact of \npEMD~on PMR has not been previously addressed, which may have contributed to a certain extent to the conflicting results for the density dependence of SE obtained using this observable. 

\begin{figure}
  \includegraphics[width=0.45\textwidth]{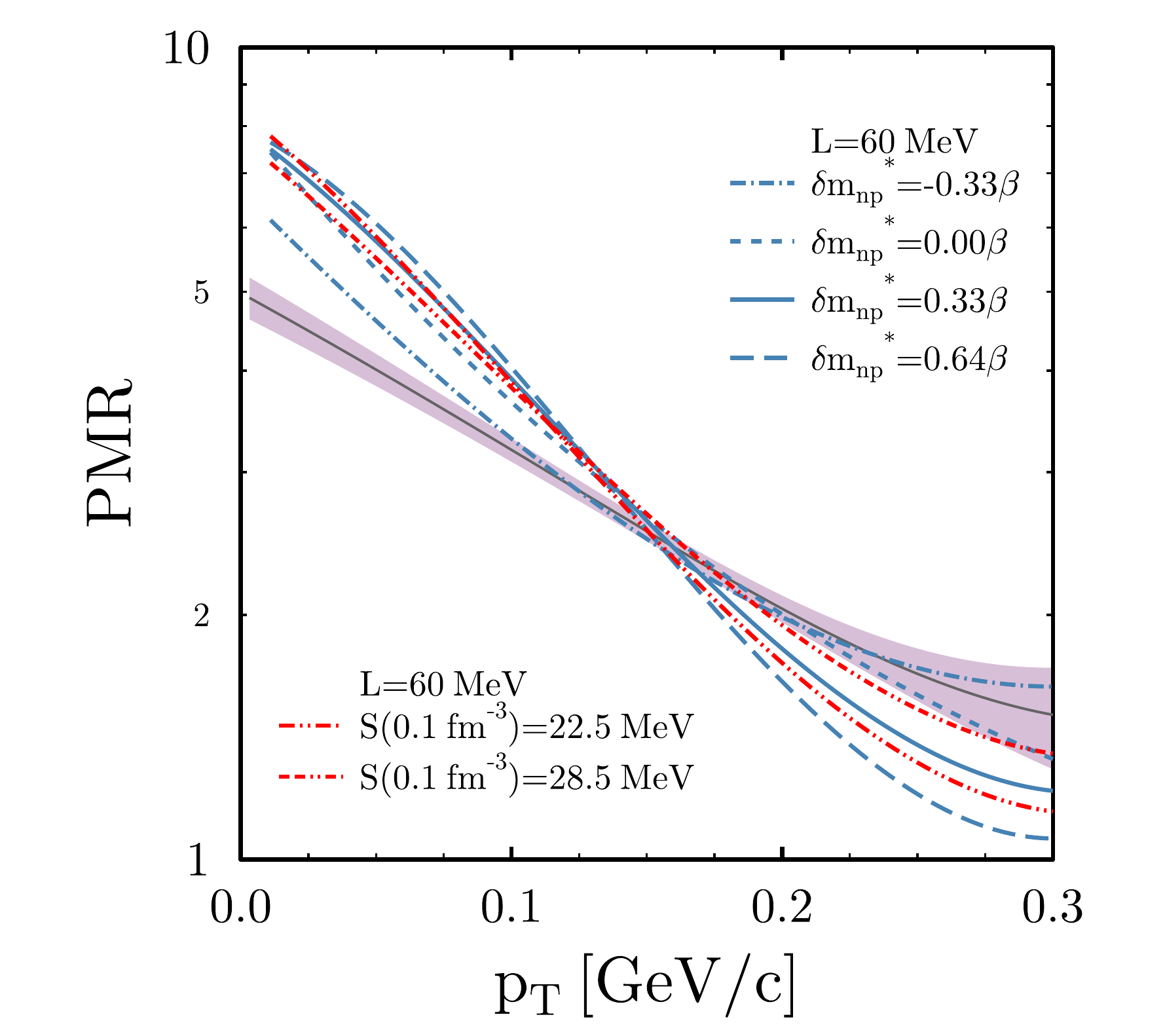}	
  \caption{Sensitivity of the $p_T$ dependent PMR spectra to the magnitude of neutron-proton effective mass difference  $\delta m_{np}^*$ and value of the symmetry energy at $\rho$=0.10 fm$^{-3}$. The same details regarding the reaction as for
  \figref{multvsavptfit} are in order.} 
  \figlab{efmisvdependence}
\end{figure}

\begin{figure}
  \includegraphics[width=0.45\textwidth]{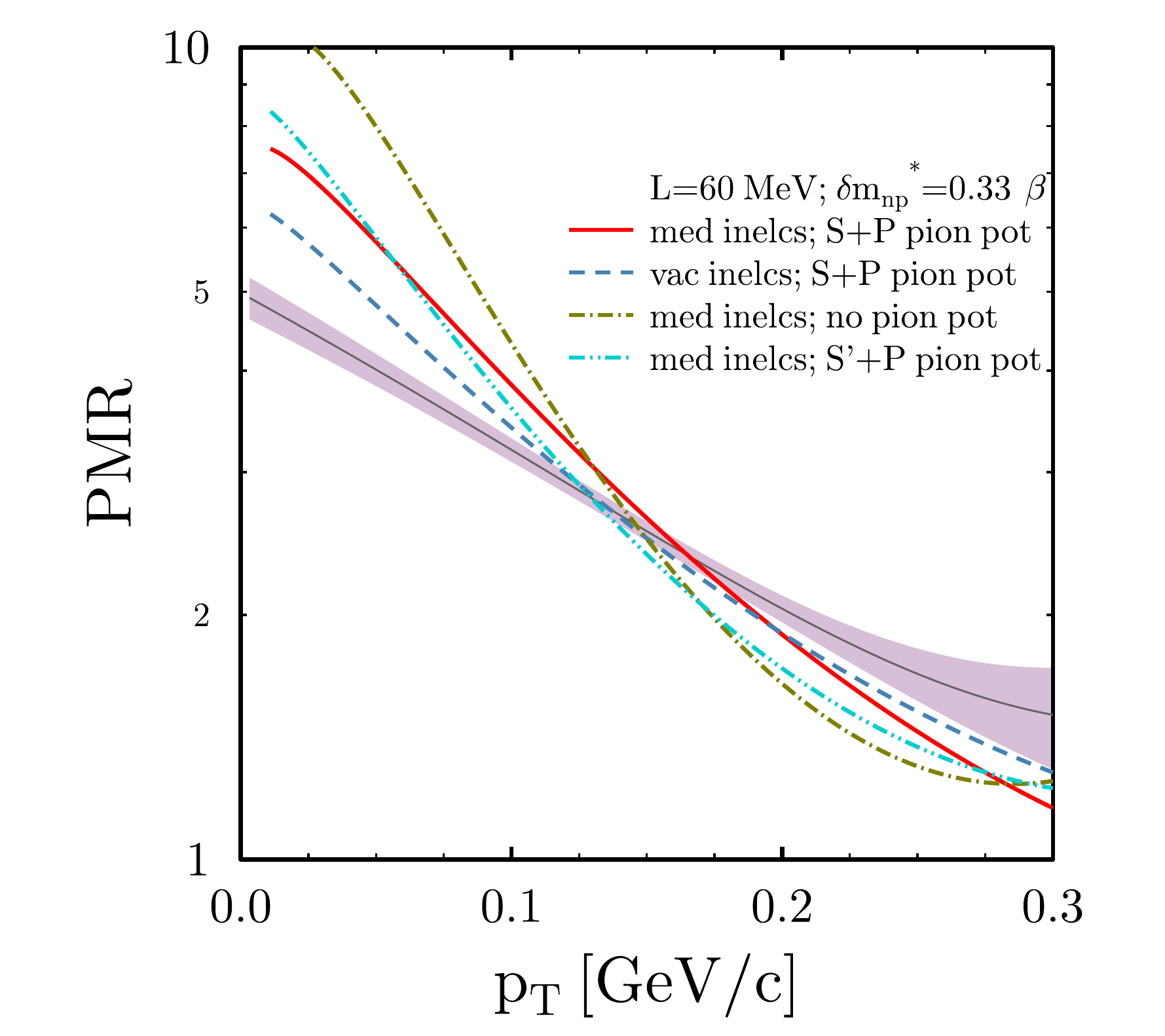}	
  \caption{Model dependence PMR spectra to the pion potential and in-medium effects on inelastic cross-sections. The same details regarding the reaction as for \figref{multvsavptfit} are in order.} 
  \figlab{pmrspectramoddep}
\end{figure}

\begin{figure*}
 \includegraphics[width=0.495\textwidth]{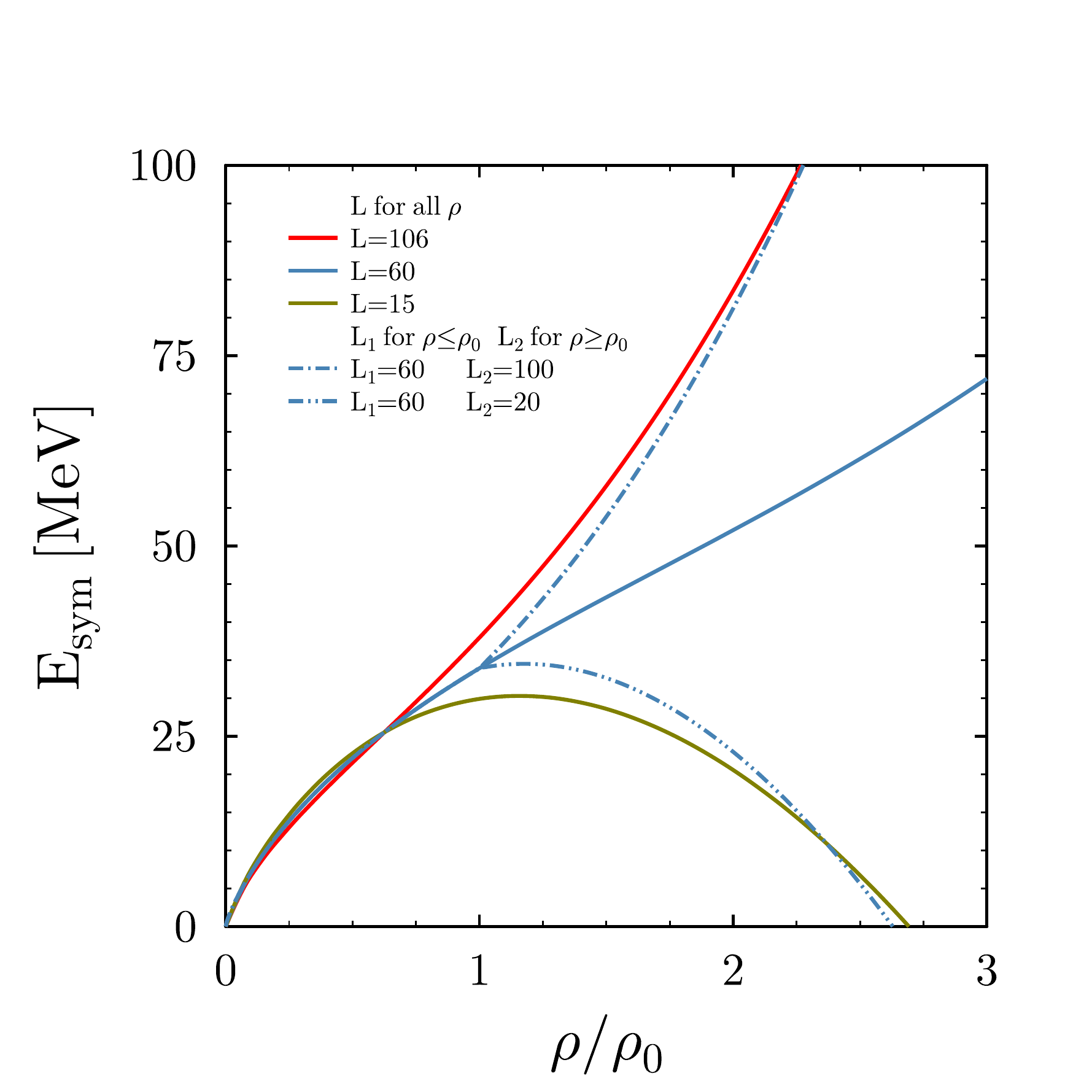}	
 \includegraphics[width=0.495\textwidth]{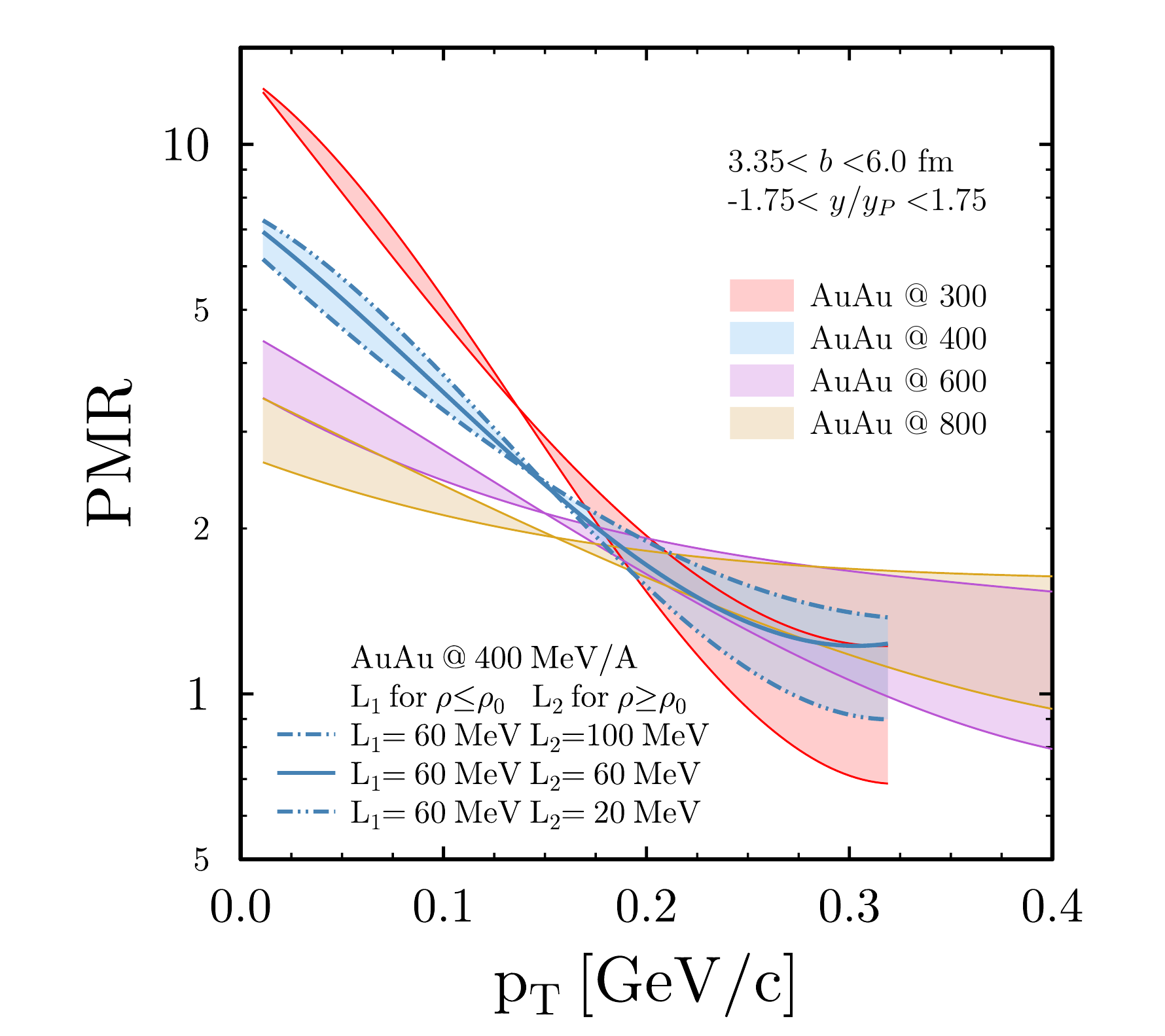}
 \caption{(Left Panel) Density dependence of the symmetry energy for the three values of $L$ for which results were
 shown in \figref{multvsavptfit}. For the $L$=60 MeV case two additional EoS'es that have the same density dependence below saturation as the standard one but have different values for the slope parameter $L$ above that point are shown. (Right Panel) Transverse momentum dependent PMR spectra for the three EoS'es with $L$=60 MeV but with different values of the slope above saturation for mid-central AuAu collisions at various impact energies. As the impact energy is increased the transverse momentum above which spectra are insensitive to residual DPOT dependence (see \figref{multvsavptfit}) also increases, requiring computation of spectra up to higher values of this variable. For AuAu collisions at 300 MeV/nucleon impact energy, the total charged pion multiplicity and ratio are determined by extrapolating existing experimental data for central AuAu reactions ~\cite{Reisdorf:2010aa} which leads to the approximate values of 1.0 and 4.25 for the two observables respectively. To obtain the corresponding result for mid-central collisions the experimentally observed fact that pion multiplicity divided by the number of participants is constant as a function of impact parameter is used.}
 \figlab{ddabovesat}
\end{figure*}

The latter parameter represents a substitute to fixing the symmetry energy at saturation, a point where it is not 
accurately known at present. It has been however possible to extract precise values at
sub-saturation densities from experimental data of static properties of nuclei \cite{Trippa:2008gr,Brown:2013mga,Zhang:2013wna}.
Such empirical findings are in good agreement with many body calculations of the neutron matter EoS that use as input microscopical N$^3$LO chiral perturbation theory effective potentials~\cite{Kruger:2013kua,Drischler:2016djf,Drischler:2015eba}.
The empirical value S($\tilde\rho$)=25.5 MeV, with $\tilde\rho$=0.1 fm$^{-3}$, extracted in Ref. \cite{Brown:2013mga} has been used as part of the standard input to the model. The sensitivity to this parameter has been studied by varying it in the extremely conservative interval 22.5 MeV $<$ S($\tilde\rho$) $<$ 28.5 MeV. Results plotted in \figref{efmisvdependence} prove that its magnitude is limited to less than 10$\%$. If the uncertainty of 1.0 MeV quoted in Ref. \cite{Brown:2013mga} for S($\tilde\rho$) is taken into account as a more realistic interval of variation, the sensitivity drops to a few percent.

The sensitivity of results to a few extra model ingredients has additionally been studied. In ~\figref{pmrspectramoddep} the impact of modifying the pion optical potential, either by choosing a different S-wave optical potential or discarding it completely, and switching off in-medium effects on inelastic cross-sections in PMR spectra is shown. In relative terms the impact of these model ingredients is largest at low $p_T$ values. Nevertheless, in the high $p_T$ region, of interest for SE studies, a 10$\%$ effect is still observed. Additionally it has been investigated what the impact on PMR spectra of setting the DPOT equal to nucleon's or to a rather arbitrary strength ($\tilde{U}_{0}^\Delta$=-25 MeV, $m_{\Delta}^*$=0.85, $\tilde{U}_{sym,1}^\Delta$= 0 MeV and $\delta m_{\Delta}^*$=-0.15) would be (not shown). In either case the deviation from the standard full model calculation in ~\figref{pmrspectramoddep} amounts to 20$\%$ in the high $p_T$ region. Fitting multiplicity observables to extract DPOT parameters is thus a minimum requirement to keep model dependence at reasonable levels.

The results presented above lead to the conclusion that studies of PMR cannot provide a constraint for the density dependence
of SE but rather a correlation of the parameter used to adjust its stiffness (here the slope $L$) with the value of \npEMD. To lift this degeneracy an independent constraint or information for the latter quantity needs to be provided
from other sources. Elliptic flow ratios of neutrons-to-charged particles, double ratio of $n/p$ multiplicity spectra and dipole polarizability of nuclei have been identified as promising such sources \cite{Cozma:2017bre,Morfouace:2019jky,Malik:2018juj}. To minimize model dependence, a third observable providing a constraint for the nucleon isoscalar effective mass may be required.

\subsection{Probed density and impact energy dependence}
\seclab{4b}
PMR ratio has been proposed as a probe of the density dependence of SE above saturation. A few studies that address this
question are available \cite{Liu:2014psa,Yong:2017hak}, but neither of these models include threshold effects. A proof that pion production probes densities significantly above saturation is provided in the following. In the left panel of \figref{ddabovesat} the density dependence of SE for the three choices of $L$ employed in this Section is presented. Two additional EoS'es that lead to different density dependence above saturation for the $L$=60 MeV case have been constructed by modifying the slope parameter above saturation to $L$=100 MeV and $L$=20 MeV to reproduce a stiff and a soft density dependence above this point respectively. The three $L$=60 MeV EoS'es have identical density dependence below saturation enforced by using in each case also a common value for the curvature parameter $K_{sym}$. 

To avoid numerical problems generated by discontinuous derivatives of the SE with respect to density, model parameters that govern the density dependence of SE become $C^1$ functions of this variable in a narrow interval around $\rho_0$, its width being set to 0.05$\rho_0$. As a consequence, additional contributions to forces proportional to the derivatives with respect to density of coupling constants will need to be included to obey energy conservation. For the two-body term in \eqref{hamiltonian} these corrections lead to computational requirements that scale with the third power of the number of nucleons, instead of the second power for the ordinary case. To avoid this issue, the coupling constant of the two-body term has been kept the same below and above saturation. Only the three-body contributions, proportional to the coupling constants $x$ and $y$ in \eqref{hamiltonian}, have been modified with the consequence that above saturation the values of $L$ and $K_{sym}$ cannot be chosen independently anymore. The advantage of this approach resides in the fact that energy conservation violation is small, of the order of a few hundred KeV per event, even without including contributions to forces due to the density dependence of the two coupling constants in the vicinity of saturation.

In the right panel of \figref{ddabovesat} theoretical values for PMR spectra are presented for the three asyEoS'es that are identical below saturation but differ above this point. Results for mid-central AuAu collisions at four impact energies in the 300-800 MeV/nucleon range are shown. Noteworthy differences between the stiffest and softest choices for the SE that amount to a factor close to 2 in the region $p_T>$ 0.25 GeV/c are observed. For the impact energy of 400 MeV/nucleon it is almost as large as the one evidenced in \figref{multvsavptfit} for the case when the asy-EoS'es also differ below saturation. This proves that the information about the high density phase where most $\Delta$(1232) are first excited is preserved to a large extent in spite of the fact that pions that survive up to the final state of the reaction undergo, on average, a few absorption/decay processes. To determine the density at which PMR is most sensitive to, calculations with different combinations of values for the slope parameter below and above saturation have to be performed. The average probed density can then be extracted using the approach described in Ref. ~\cite{Morfouace:2019jky}. The sensitivity to the asy-EoS above saturation is approximately independent on impact energy advocating experimental measurements at higher impact energy in view of less required beam-time for similar statistical accuracy.

\section{Summary and Conclusions}
\seclab{5}

The dcQMD model, an offspring of the T\"ubingen QMD transport model, has been further developed by implementing in-medium nucleonic resonance potentials that can be set independently of the nucleon optical potential and are described in terms of intuitive quantities such as potential depths, at saturation and zero momentum, and effective masses. This effort has been prompted by results of phenomenological studies and ab-initio calculations that suggest a $\Delta$ (1232) potential that is different from that of the nucleon. The two approaches have led however to different results which has contributed in the past to adopting the Ansatz of equal resonance and nucleon potentials in semi-classical transport models for heavy-ion collisions at intermediate impact energies. This model extension has been deemed necessary as the accurate description of observables carrying information about the isospin dependent part of the equation of state of nuclear matter requires a proper understanding of residual effects induced by uncertainties of our knowledge of the equation of state of symmetric nuclear matter or other quantities leading to isoscalar contributions to observables.

The upgraded model has been employed in the study of pion production from slightly above threshold to impact energies of 800 MeV/nucleon. One of the objectives has been the extraction of ${\it effective}$ isoscalar and isovector $\Delta$(1232) potential strengths and masses from a comparison to available experimental data for $^{40}$Ca$^{40}$Ca, $^{96}$Ru$^{96}$Ru, $^{96}$Zr$^{96}$Zr and $^{197}$Au$^{197}$Au provided by the FOPI Collaboration. The analysis has been performed separately for the isoscalar and isovector components of the $\Delta$(1232) potential following the identification of observables that are predominantly sensitive to one of the two potentials: total charged pion multiplicity for the former and ratio of total charged multiplicity for systems with different isospin asymmetry for the latter. The charged pion multiplicity, an observable proposed in the past for the study of the density dependence of symmetry energy, has been shown to be equally sensitive to both the isoscalar and isovector $\Delta$(1232) potentials. It has been shown that available experimental data for nucleonic observables such as stopping, transverse and elliptic flow for systems of different masses and at different impact energies can be accurately described by the model, a pre-requisite for studying pion emission close to threshold.

The extraction of the isoscalar $\Delta$(1232) potential (ISDP) parameters has been attempted using the experimental data for total charged pion multiplicities for $^{40}$Ca$^{40}$Ca and also $^{96}$Ru$^{96}$Ru systems at impact energies of 400, 600 and 800 MeV/nucleon (only the first impact energy for the latter system). A precise extraction of the potential depth and isoscalar effective mass was not possible due to sub-optimal accuracy of experimental data and a decrease of sensitivity at higher impact energies. However, an effective isoscalar potential that is more attractive and a smaller isoscalar effective mass are favored for $\Delta$(1232) as compared to those corresponding to the nucleon. The result is in agreement with similar information extracted from quasi-elastic electron-nucleus scattering but is incompatible with microscopical model calculation. A possible reason for the latter is the omission of non-resonant pion production contributions, which would lead to a less attractive potential and may also impact its required momentum dependence.

For the isovector $\Delta$(1232) potential (IVDP) the study has proven more challenging. Comparing model predictions for the ratio of total charged multiplicity for systems with different isospin asymmetry to experiment has only led to extremely loose constraints for the IVDP parameters, spanning half of the probed parameter space. Using the pion multiplicity ratio for isospin asymmetric systems instead, more precise constraints, in the form of correlations between potential depth and isovector effective mass difference, could be obtained. These have however proven to be rather sensitive to values of the slope parameter of symmetry energy at saturation, the value of the neutron-proton effective mass difference and the assumed stiffness for the density dependence of IVDP above saturation. Adding the total charged pion multiplicity to the fit was shown to exclude the more repulsive IVDP scenarios.

Without an accurate knowledge of the $\Delta$(1232) potential a study of the symmetry energy using multiplicity observables alone is not possible.  An alternative, previously studied in Ref. \cite{Cozma:2016qej}, is to include average transverse momentum observables among the fitted quantities. The additional uncertainties induced by the ISDP results however in even more uncertain constraints than before. 

Studying pion multiplicity ratio spectra has proven more fruitful. It has been shown that by including average transverse momenta in the fit of DPOT parameters, a value of $p_T$ above which spectra are insensitive to uncertainties in the $\Delta$(1232) potential can be determined. Residual model dependence due to pion optical potential and in-medium modifications of cross-sections uncertainties are below 10$\%$ in this high $p_T$ region. Extraction of information regarding the symmetry energy and related quantities is thus feasible from a comparison theory-experiment of the high $p_T$ tail of pion multiplicity ratio spectra. It has been shown that due to inclusion of threshold effects the sensitivity to the value of the neutron-proton effective mass difference has to be taken into consideration. Without input from other sources only a correlation between the values of the slope of symmetry energy and neutron-proton effective mass can be extracted from pion production close to threshold. The sensitivity to the magnitude of symmetry energy at $\rho$=0.10 fm$^{-3}$, the density for which it is kept fixed in the present model, was found to be small, of the order of a few percent. The sensitivity on the density dependence of the symmetry energy above saturation was however found appreciable in spite of the fact that surviving pions undergo, on average, a few absorption/decay cycles and was proved to be approximately independent of impact energy.

It is concluded that more accurate experimental data for pion production in heavy-ion collisions from threshold to 800 MeV/nucleon incident energy, that provide sufficient statistical accuracy but are below the point where the fraction of excited nucleons into resonances becomes non negligible, will be one of the pre-requisites for the extraction of constraints for the symmetry energy at supranormal densities from terrestrial laboratory experiments. However, precise information from other sources regarding the momentum dependence of the isovector component of the nucleon potential will be needed for providing a precise answer regarding the value of the symmetry energy around 2$\rho_0$.

\section{Acknowledgments}
The authors acknowledge financial support from the U.S. Department of Energy, 
USA under Grant Nos. DE-SC00145 30, US National 
Science Foundation, United States Grant No. PHY-
1565546. The research of M.D.C. has been financially supported in part by the Romanian Ministry of Education and Research through Contract No. PN 19 06 01 01/2019-2022. M.D.C. acknowledges the hospitality of NSCL / MSU where part of this study was performed. The authors express their gratitude to Maria Colonna, Pawel Danielewicz, Justin Estee, Che-Ming Ko, William Lynch, Hermann Wolter and TMEP Collaboration for stimulating discussions. The computing resources have been partly provided by the Institute for Cyber-Enabled Research (ICER) at Michigan State University.

%% For one-column wide figures use
%\begin{figure}
%% Use the relevant command to insert your figure file.
%% For example, with the graphicx package use
%  \includegraphics{example.eps}
%\end{figure}
%%
%% For two-column wide figures use
%\begin{figure*}
%% Use the relevant command to insert your figure file.
%% For example, with the graphicx package use
%  \includegraphics[width=0.75\textwidth]{example.eps}
%\end{figure*}
%%

%\begin{table}
%\begin{tabular}{lll}
%\hline\noalign{\smallskip}
%\end{tabular}
%\end{table}

%\begin{acknowledgements}
%\end{acknowledgements}

\bibliographystyle{spphys}       % APS-like style for physics
\bibliography{references}   % name your BibTeX data base

\end{document}